\documentclass[pre,twocolumn,superscriptaddress,showpacs]{revtex4}
\usepackage{amsmath}
\usepackage{amssymb}
\usepackage{graphicx}

\newcommand{\T}{${\mathcal T}\,$}
\newcommand{\Ti}{${\mathcal T}$}

\begin{document}

\title{Quantum Chaotic Scattering in Microwave Resonators}

\author{B.~Dietz}
\affiliation{Institut f{\"u}r Kernphysik, Technische Universit{\"a}t
Darmstadt, D-64289 Darmstadt, Germany}

\author{T.~Friedrich}
\affiliation{Institut f{\"u}r Kernphysik, Technische Universit{\"a}t
Darmstadt, D-64289 Darmstadt, Germany}
\affiliation{GSI Helmholzzentrum f{\"u}r Schwerionenforschung GmbH, D-64291
Darmstadt, Germany}

\author{H.~L.~Harney}
\affiliation{Max-Planck-Institut f{\"u}r Kernphysik, D-69029 Heidelberg,
Germany}

\author{M.~Miski-Oglu}
\affiliation{Institut f{\"u}r Kernphysik, Technische Universit{\"a}t
Darmstadt, D-64289 Darmstadt, Germany}

\author{A.~Richter}
\email{richter@ikp.tu-darmstadt.de}
\affiliation{Institut f{\"u}r Kernphysik, Technische Universit{\"a}t
Darmstadt, D-64289 Darmstadt, Germany}
\affiliation{$\rm ECT^*$, Villa Tambosi, I-38100 Villazzano (Trento), Italy}

\author{F.~Sch{\"a}fer}
\affiliation{Institut f{\"u}r Kernphysik, Technische Universit{\"a}t
Darmstadt, D-64289 Darmstadt, Germany}

\author{H.~A.~Weidenm{\"u}ller}
\affiliation{Max-Planck-Institut f{\"u}r Kernphysik, D-69029 Heidelberg,
Germany}

\date{\today}

\begin{abstract}
In a frequency range where a microwave resonator simulates a chaotic
quantum billiard, we have measured moduli and phases of reflection and
transmission amplitudes in the regimes of both isolated and of weakly
overlapping resonances and for resonators with and without
time-reversal invariance. Statistical measures for $S$-matrix
fluctuations were determined from the data and compared with extant
and/or newly derived theoretical results obtained from the
random-matrix approach to quantum chaotic scattering. The latter
contained a small number of fit parameters. The large data sets taken
made it possible to test the theoretical expressions with
unprecedented accuracy. The theory is confirmed by both, a
goodness-of-fit-test and the agreement of predicted values for those
statistical measures that were not used for the fits, with the data.
\end{abstract}


\maketitle


\section*{Introduction}
\label{sec:intro}

Microwave resonators, also known as ``microwave billiards'', are ideal
systems to study properties of chaotic quantum
systems~\cite{Sto90,Sri91,Gra92,mw:chaos,richter:playing}. Most
studies have focused on the statistical properties of eigenvalues and
eigenfunctions, especially on tests of the
Bohigas-Giannoni-Schmit~\cite{BGS,Boh84,heusler,QC} conjecture.
According to that conjecture, the spectral fluctuation properties of
quantum systems with chaotic classical dynamics coincide with those of
random-matrix ensembles belonging to the same symmetry class. That
statement holds up to level spacings determined by the period of the
shortest periodic orbit of the system. We comment on that point below.
Data are taken by coupling the resonators via one or several antennas
to sources or sinks of a microwave power supply. Because of this
arrangement, microwave resonators can also be viewed as open quantum
systems, and measurements of the reflected and transmitted intensity
amplitudes provide generic information on chaotic quantum scattering,
each antenna acting as a single scattering channel~\cite{scattering}.

In this paper we report on measurements of the complex transmission
and reflection amplitudes of chaotic microwave billiards, and on the
theoretical analysis of such data. For the latter we use the generic
approach to chaotic quantum scattering based on random-matrix theory
(RMT). We have used two types of microwave resonators. In the first
one, time-reversal (\Ti) invariance holds and in the second one, it is
violated by placing a magnetized ferrite within the cavity. The large
set of scattering data taken with either device (considerably larger
than data sets collected, for instance, in nuclear physics) allows us
to test the RMT approach to chaotic scattering with unprecedented
accuracy, both for systems that are \T invariant and for those that
are not. Most experimental investigations of chaotic scattering have
been restricted to measure cross sections rather than individual
elements of the scattering matrix $S$. In our setup we use a vector
network analyzer to actually measure modulus and phase of the
reflected and of the transmitted amplitudes and, thus, of individual
$S$-matrix elements. The additional information garnered in this way
increases the significance of our tests. In addition to testing the
RMT approach to chaotic scattering, we propose and test a method for
the determination of the strength of \T violation from data on
$S$-matrix correlation functions. This is of particular significance
for those chaotic quantum scattering systems for which the relevant
parameters cannot be determined easily by dynamical calculations such
as the semiclassical approximation \cite{Bluemel1998}. Some of our
results have already briefly been reported in
Refs.~\cite{Friedrich2008,Schaefer2009}.

For \Ti-invariant systems statistical cross-section fluctuations have
been thoroughly investigated experimentally and compared with
theoretical predictions in the regime of isolated nuclear
resonances~\cite{Lyn68} (average resonance spacing $d$ very large
compared to average resonance width $\boldsymbol\Gamma$) and in the Ericson
regime~\cite{Ericson} ($\boldsymbol\Gamma \gg d$), especially in
nuclei~\cite{Eri63}, but also in several other
systems~\cite{Blu88,Lawniczak2008,Schaefer2003}. We are not aware of
similarly extended and precise tests of the RMT approach to chaotic
scattering in the regime of weakly overlapping resonances ($\boldsymbol\Gamma
\sim d$). Our work is intended to fill that gap. \Ti-invariance
violation was tested in nuclear spectra~\cite{Fre85} and for the
Ericson regime in compound-nuclear
reactions~\cite{Ericson:66,mahaux:66,Witsch1967,Boose,Harney90}. Upper
bounds on the strength of the \Ti-invariance-violating interaction
were deduced in both cases. \Ti-invariance violation caused by an
external magnetic field has also been studied in electron transport
through quantum dots~\cite{Pluhar1995} and other
devices~\cite{Bergman} and in ultrasound transmission in rotational
flows~\cite{Rosny}. The RMT approach to \Ti-invariance
violation~\cite{Boose,Pluhar1995} used in some of these papers is
likewise tested very precisely in the present paper.

The theoretical approach to chaotic scattering is based on an
expression for the $S$-matrix originally derived in the context of
nuclear physics~\cite{mahaux:69}. That expression contains explicitly
the Hamiltonian matrix of the system. Replacing the actual Hamiltonian
by a \Ti-invariant random-matrix ensemble, one generates an ensemble
of $S$-matrices which describes generic features of chaotic
scattering. Analytical expressions for the $S$-matrix correlation
functions of that ensemble which apply for all values of $\boldsymbol\Gamma / d$
have been derived~\cite{Ver85}. These are used in our analysis.
Replacing the Hamiltonian by an ensemble of random matrices with
partially broken \T invariance~\cite{Pandey}, one similarly generates
an ensemble of $S$-matrices that describes generic features of chaotic
scattering with broken \T invariance. Some properties of that ensemble
have been worked out previously~\cite{Pluhar1995,Fyodorov2005}. To
compare with our data we had to extend the theoretical results. This
work is also reported in the present paper.

The paper is organized as follows. In Sec.~\ref{sec:expt} we describe
the experimental setup and some typical results. In
Sec.~\ref{sec:measures} we define the statistical measures in terms of
$S$-matrix elements and use these to analyze the data. In particular,
we define a measure that quantifies the strength of \Ti-invariance
violation. In Sec.~\ref{sec:theory} and in the Appendix we sketch the
derivation of analytic expressions for the statistical measures. We
use the method of Ref.~\cite{Pluhar1995}. The theory contains a number
of parameters. These are fitted to data. We test the theory with the
help of a goodness-of-fit (GOF) test in Sec.~\ref{sec:gof}. The basic
assumption for the applicability of the GOF test is that the
distribution of the Fourier-transformed $S$-matrix elements are
Gaussian and uncorrelated. In Sec.~\ref{sec:FTdistrib} we demonstrate
the validity of that assumption for chaotic scattering systems. In the
case of \Ti-invariance violation, we test the theory further by
comparing experimental values for the elastic enhancement factor and
for the distribution of the diagonal $S$-matrix elements with
theoretical predictions based on parameter fits to other observables.

\section{Experiment}
\label{sec:expt}

\begin{figure}[ht]
	\includegraphics[width=8cm]{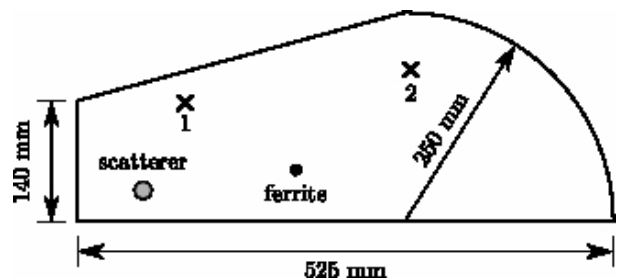} 
	\caption{The tilted stadium billiard (schematic). The two antennas
	$1$, $2$ connect the resonator to the VNA. Optionally a ferrite is
	inserted at a fixed location to violate \T invariance and/or a
	movable scatterer is used to gather independent data sets (see main
	text). Taken from Ref.~\cite{Schaefer2009}.}
	\label{fig:tilted_stadium}
\end{figure}

For an experimental study of universal fluctuation properties of
chaotic scattering systems we used flat, cylindrical microwave
resonators with the emitting and receiving antennas acting as single
scattering channels. As long as the excitation frequency $f$ is chosen
below $f_{\rm max} = c_0/(2\,h)$, where $h$ is the height of the
resonator and $c_0$ is the speed of light, only transverse magnetic
TM$_0$ modes can be excited, and the electrical field vector is
perpendicular to the top and the bottom plates of the resonator. Then,
the associated Helmholtz equation for the electric field strength is
scalar and mathematically identical to the two-dimensional
Schr\"odinger equation of a particle elastically reflected by the
contour of the microwave resonator, i.e., of a quantum
billiard~\cite{Sto90}. The experiments were performed with resonators
whose contour has the shape of a tilted stadium
billiard~\cite{Primack1994}. That shape was chosen to avoid
bouncing-ball orbits. Each microwave resonator was constructed from
three metallic plates. The bottom and the top of the resonator
are formed by two 5~mm thick high-purity copper plates. The center
plate had a hole in the shape of a tilted quarter stadium. The
thickness of that plate determined the height of the resonator and
differed for the two experiments. The quality factor Q increases with
the height of the resonator. Thus, to ensure a high Q value the center plate
for the \Ti-invariant case had a thickness of 14.6~mm (so that
$f_{\rm max} = 10.3~{\rm GHz}$). The plate was made of
brass~\cite{Friedrich2008} to technically permit the cutting-out
of the hole. For the case with broken \T invariance a copper plate
with a thickness of 5.0~mm, which coincided with the height of the 
ferrite described below, was used (so that $f_{\rm max} = 30.0~{\rm
GHz}$), see~Ref.~\cite{Schaefer2009}. In
order to make sure that only TM$_0$ modes are excited, the excitation
frequency $f$ was actually chosen $\leq 25~{\rm GHz}$. Screws through
the top, middle and bottom plates ensured the good electrical contact
needed to achieve high-quality values of the resonator. Two thin wires
(diameter about 0.5~mm) intrude 2.5~mm into the cavity through
small holes (diameter about 2~mm) drilled into the lid of the
resonator. They act as dipole antennas to couple the rf power into and
out of the resonator. A vector network analyzer (VNA) provided the rf
signal at a variable frequency $f$ and recorded the signal received at
the same (other) antenna for reflection (transmission)
measurements. The two signals were compared by the VNA in amplitude
and phase to determine the complex-valued $S$-matrix elements. These
formed the data set for our analysis. The cavity is schematically
shown in Fig.~\ref{fig:tilted_stadium}.

\begin{figure}[ht]
	\includegraphics[width=8cm]{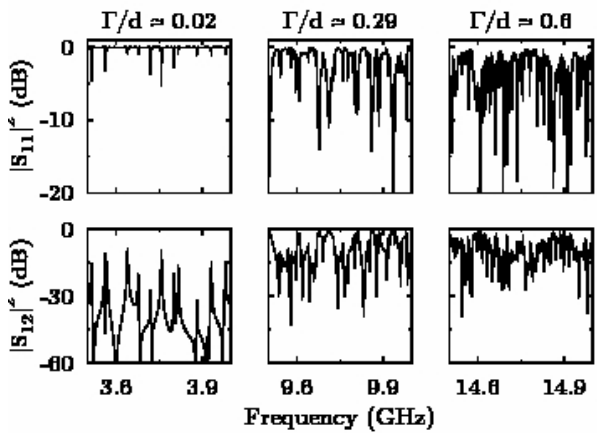}
	\caption{Reflection spectra (upper panels) and transmission
	spectra (lower panels) of the \Ti-invariant billiard taken at
	three frequency ranges (panel columns). While the left and
	center panels show data at $\Gamma / d \approx 0.02$ and
	$0.29$, respectively, for the two-dimensional regime (where
	the billiard mimicks a quantum billiard), the data at $\Gamma
	/ d \approx 0.6$ in the right panels are obtained in a
	frequency range where the cavity supports three-dimensional
	field distributions.}
\label{fig:goe_spectra}
\end{figure}

For a precise experimental determination of the elements of the
$S$-matrix all systematic and statistical errors must be minimized.
The coaxial lines connecting the VNA with the cavity are the dominant
source for systematic errors. They attenuate and reflect the rf
signal. Both effects were removed by a proper calibration of the VNA.
Systematic errors are also caused by the transmission properties
of the two antennas. To account for these, the reflection spectrum of
a small cylindrical resonator (diameter 5~mm, depth 20~mm) was
measured using the same antenna geometry as in the actual
experiment. The first resonance is located well above 30~GHz. Thus,
in the frequency range of interest and for an ideal coupling of the
antennas to the resonator all rf power would be reflected. Any
deviation from this expectation was attributed to the antennas. The
resulting correction was applied to the measured spectra in the actual
experiments. The corrected values of the reflection and transmission
spectra with the two antennas $1$ and $2$ provided the elements $S_{1
1}, S_{1 2}, S_{2 1}$ and $S_{2 2}$ of the complex $2\times2$
$S$-matrix as functions of the frequency $f$. The frequency step size
$\Delta f$ was $\geq 100~{\rm kHz}$.  Typical measured reflection and
transmission spectra are shown in Fig.~\ref{fig:goe_spectra}. For the
measurement of the $S$-matrix element $S_{11}(f)$ antenna 1 was used
as emitting and receiving antenna, for that of $S_{12}(f)$ antenna 2
was used as emitting, antenna 1 as receiving antenna,
etc.. Figure~\ref{fig:goe_spectra} shows that at low excitation
frequencies the resonances of the billiard are isolated, i.e.\ the
mean resonance width $\boldsymbol\Gamma$ and the correlation width 
$\Gamma$, is small compared to the mean level spacing $d$. 
Since it is a difficult if not impossible task to determine the 
resonance widths in the regime of overlapping resonances whereas the 
correlation width $\Gamma$ can be well estimated from 
the data using the Weisskopf formula~\cite{Bla52} given in Eq.~(\ref{eqn:GammaD}) below 
(for more details see Sec.~\ref{fitpar}), we refer to the latter in the following.

As $f$ increases, so does 
the ratio $\Gamma / d$, and the
resonances begin to overlap.

The statistical errors of a single measurement caused by thermal
fluctuations were reduced by an internal averaging routine of the VNA.
The resulting errors were several orders of magnitude smaller than the
signal and, thus, negligible. The data were analyzed in frequency
intervals of $1$ GHz length yielding $M\approx 10^4$ data points each.
The limited number of statistically independent data points in every
such frequency interval causes finite-range-of-data (FRD) errors. To
increase the number of data points and to reduce the FRD
errors~\cite{Gibbs1965, Dallimore1966}, in some of the experiments a
small scatterer (an iron disc, 20~{\rm mm} diameter) was introduced
into the microwave resonator (see Fig.~\ref{fig:tilted_stadium}) and
moved to six different positions. We then speak of different
\emph{realizations} of the scattering system.

Experiments with violated \T invariance were done with a magnetized
ferrite embedded within the resonator. Such \emph{induced}
Time-Reversal-Invariance Violation (TRIV) has been studied in numerous
works~\cite{So1995, Stoffregen1995, Wu1998, Hul2004, Schaefer2007,
Schaefer2009,Vranicar}. The ferrite has a cylindrical shape (4~mm
diameter, 5~mm height), a saturation magnetization $4\pi\, M_{\rm S} =
1859~{\rm Oe}$, and a linewidth $\Delta H = 17.5~{\rm Oe}$, with
1~Oe=1000/4$\pi$~A/m. It was provided by courtesy of AFT Materials
GmbH (Backnang, Germany). Two NdFeB magnets (cylindrical shape, 20~mm
diameter and 10~mm height) were placed outside the billiard at the
position of the ferrite to provide the required magnetic fields
perpendicular to the top and bottom plates of the resonator. The
distance between the magnets and the ferrite could be adjusted by a
screw thread mechanism, and field strengths of up to 360~mT could be
achieved at the position of the ferrite. With this setup TRIV is
induced via the following mechanism.  Because of the external magnetic
field ${\bf B}$ the ferrite effectively acquires a macroscopic
magnetization ${\bf M}$ that precesses with the Larmor frequency
$\omega_0$ around ${\bf B}$. This is the origin of the
\emph{ferromagnetic resonance}. The rf magnetic field inside the
cavity is elliptically polarized and can be decomposed into two field
components of opposite circular polarization with, in general,
different amplitudes. Due to the Larmor precession of the
magnetization ${\bf M}$ the spins of the ferrite couple differently to
the two magnetic field components. A reversal of time, simulated by an
interchange of the input and output channels, swaps the rotational
sense of the two field components and thus, due to their different
amplitudes, effectively changes the coupling of the ferrite to the
resonator mode. The induced TRIV is strongest if the frequency $f$ is
close to that of the ferromagnetic resonance. The experiments with the
embedded ferrite demanded a reduction of the height of the resonator
to 5.0~mm, as the ferrite itself was only 5~mm in height.

\begin{figure}[b]
	\includegraphics[width=8cm]{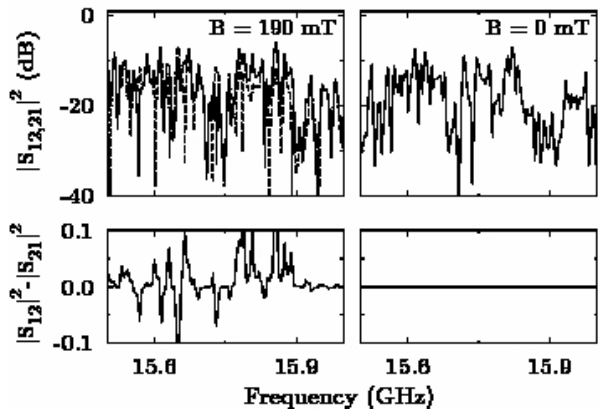} \caption{Spectra
	of the tilted stadium billiard with a TRIV ferrite magnetized
	through an external field of $B=190~{\rm mT}$ (left panels) and
	of $B=0~{\rm mT}$ (right panels). The upper two panels show the 
	squared modulus of the elements of the
	scattering matrix $S_{12}$ (full line) and $S_{21}$ (dashed line) 
	taken in the range 15.5--16.0~GHz, where $\Gamma/d \approx 0.50$. In
	this range the resonator supports, due to its height of 5~mm,
	only two-dimensional modes. Reciprocity is violated for nonvanishing
	magnetic field since $S_{12} \neq S_{21}$. To clarify this, we show in 
	the lower panels the difference of the squared modulus of the elements of the
        scattering matrix $S_{12}$ and $S_{21}$.}  
	\label{fig:gue_spectra}
\end{figure}

In a \Ti-invariant system, the scattering matrix is symmetric, $S_{1
2} = S_{2 1}$. We refer to that property as reciprocity. Violation of
reciprocity is the hallmark of TRIV. In the setup without ferrite
the transmission spectrum for $S_{21}(f)$ is
indistinguishable from that for $S_{12}(f)$ and
reciprocity holds within the limits given by thermal noise (see left 
panels of Fig.~\ref{fig:gue_spectra}). Typical
transmission spectra of the billiard with ferrite and an external
magnetic field of $B=190~{\rm mT}$ are shown in the right panel
Fig.~\ref{fig:gue_spectra}. The two graphs in the upper panels correspond to
$|S_{12}|^2$ and $|S_{21}|^2$, in the lower panel their difference is shown.
Figure~\ref{fig:gue_spectra} demonstrates that reciprocity is
violated.

\section{Statistical measures}
\label{sec:measures}

In the present Section we define the statistical measures and use them
to analyze the data.  As pointed out in the Introduction, it is our
aim to use the data for a detailed and accurate test of random-matrix
theory. Our measures are tailored to this objective. They do not
address properties of individual resonances but instead correlation
properties of the fluctuating part of $S$-matrix elements $S_{a b}(f)$
where $a$ and $b$ take either of the values $1$ and $2$.

\subsection{$S$-Matrix Correlation Functions}

We decompose the frequency-dependent $S$-matrix into an average and
a fluctuating part,
\begin{equation}
	S_{a b}(f) = \langle S_{a b} \rangle + S^{\rm fl}_{a b}(f) \ .
	\label{eqn:Sflab}
\end{equation}
Here and in what follows, the angular brackets $\langle \ldots
\rangle$ denote an average over a suitable frequency interval. In order
to ensure a more or less constant coupling of the electric field modes
to the antennas and to the walls of the resonator we have always used
intervals of $1$ GHz length.

\begin{figure}[t]
	\includegraphics[width=8cm]{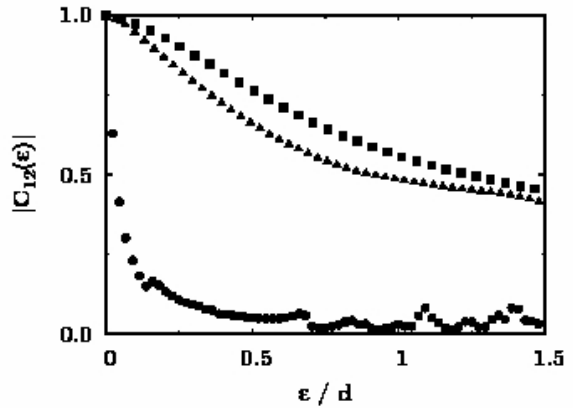} \caption{Three
	autocorrelation functions for the \Ti-invariant billiard
	determined from the measured values for $S_{12}$ in the
	frequency intervals 3--4~GHz ($\Gamma/d \approx 0.02$,
	circles), 9--10~GHz ($\Gamma/d \approx 0.29$, triangles) and
	14--15~GHz ($\Gamma/d \approx 0.6$, squares). The frequency
	difference $\varepsilon$ is plotted in units of the local mean
	level spacing $d$ as obtained from the Weyl
	formula~\cite{Bal76}. All curves are normalized to unity at
	$\varepsilon = 0$.}  \label{fig:goe_correl}
\end{figure}

The autocorrelation function of $S_{ab}(f)$ is defined by
\begin{eqnarray}
	C_{ab}(\varepsilon) &=& \langle
	S_{ab}(f)\,S_{ab}^\ast(f+\varepsilon) \rangle - \vert \langle
	S_{ab}(f) \rangle \vert^2\, \nonumber \\
	&=& \langle S_{ab}^{\rm fl}(f)\,{S_{ab}^{\rm
	fl}}^\ast(f+\varepsilon) \rangle\,.
\label{eqn:1}
\end{eqnarray}
This function quantifies the correlation between $S_{ab}^{\rm fl}$ and
${S_{ab}^{\rm fl}}^\ast$ at two different frequencies $f$ and
$f+\varepsilon$. Figure~\ref{fig:goe_correl} shows three examples of
autocorrelation functions all obtained from data for the billiard
without ferrite. The rate of decrease of the functions with increasing
$\varepsilon$ depends on the ratio $\Gamma / d$. None of the functions
has the Lorentzian shape predicted by Ericson~\cite{Ericson} for the
regime of strongly overlapping resonances $\Gamma \gg d$. We show
later that the rate of decrease agrees with random-matrix predictions
for the relevant values of $\Gamma / d$.

To quantify TRIV we measure the violation of reciprocity by the
cross-correlation function of $S_{12}(f)$ and $S_{21}^\ast(f)$,
\begin{equation}
	C_{\rm cross}(\varepsilon) = \frac{\mathfrak{Re}\left(\langle
	S_{12}^{\rm fl}(f) \, {S_{21}^{\rm
	fl}}^\ast(f+\varepsilon)\rangle\right)} {\sqrt{ \langle
	|S_{12}^{\rm fl}(f)|^2\rangle\, \langle |S_{21}^{\rm
	fl}(f)|^2\rangle}} \, .  \label{eqn:cross}
\end{equation}
For a \Ti-invariant system, reciprocity holds, and $C_{\rm cross}(0)
= 1$. In case of complete TRIV we expect that $S_{1 2}$ and $S_{2 1}$
are completely uncorrelated, $C_{\rm cross}(\varepsilon) = 0$ for all
values of $\varepsilon$. (This expectation is borne out in
Sec.~\ref{sec:theory}, see also Ref.~\cite{Schaefer2007} for a
treatment of the two-level case). In summary we have
\begin{equation}
	C_{\rm cross}(0) = \left\{
	\begin{array}{lll}
		1 & \mbox{for \T invariance \ ,} \\
		0 & \mbox{for complete TRIV \ .}
	\end{array}
	\right.
	\label{eqn:crosslimit}
\end{equation}

\begin{figure}[t]
	\includegraphics[width=8cm]{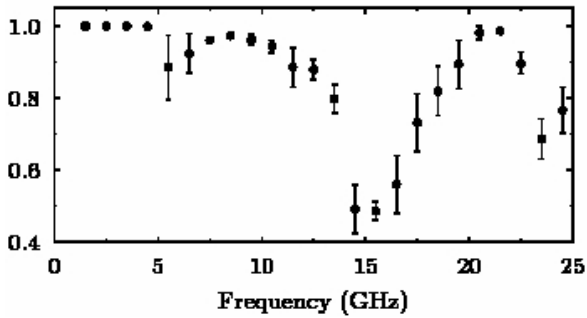}
	\caption{Cross-correlation coefficient $C_{\rm cross}(0)$ as
	a measure of TRIV. In each frequency interval of $1$ GHz
	length $C_{\rm cross}(0)$ was evaluated as an average over six
	realizations. The points denote the mean values and the error
	bars denote the standard deviations. The zero on the ordinate
	is suppressed. Based on Ref.~\cite{Schaefer2009}.}
	\label{fig:crosscorrel}
\end{figure}

As explained in Sec.~\ref{sec:expt}, we have used six realizations to
increase the statistical significance of the data. For each
realization the cross-correlation coefficient $C_{\rm cross}(0)$ was
computed and the average over all realizations was taken. For an
external magnetic field of 190~mT the resulting averaged
cross-correlation coefficient is shown in Fig.~\ref{fig:crosscorrel}.
This coefficient deviates noticeably from unity around 6, 16 and
24~GHz, indicating TRIV. The first dip can be attributed to the
ferromagnetic resonance which in our case is located at 6.6~GHz. We
assume that the other two dips arise from an enhancement of the
influence of the ferromagnetic resonance by standing rf magnetic
fields inside the ferrite. Yet, the smallest values ($C_{\rm
cross}(0) \approx 0.4$) obtained are well above zero. Hence, at a
field strength of 190~mT the ferrite induces only a partial violation
of \T invariance. This is also found at the other investigated field
strengths up to 340~mT. In Sec.~\ref{sec:theory} we show that the
strength of TRIV can be deduced from $C_{\rm cross}(0)$.

\subsection{Fourier Transformation}
\label{fourtrafo}

\begin{figure}[b]
	\includegraphics[width=8cm]{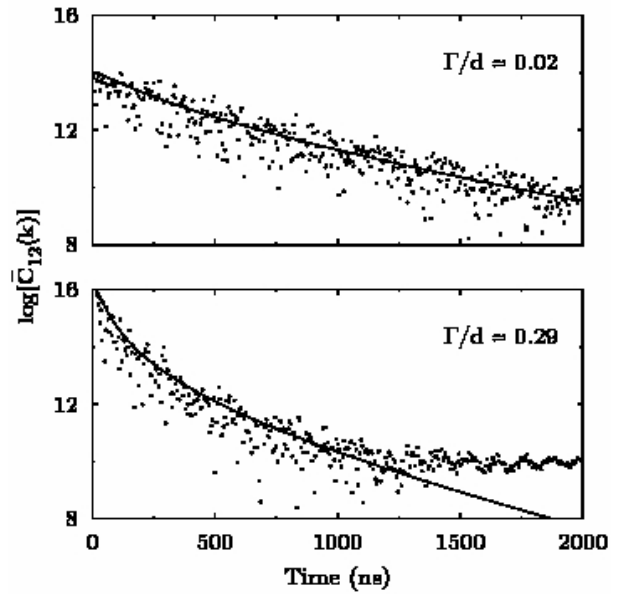} \caption{Fourier
	coefficients of the autocorrelation function of $S_{12}$ (in
	semi-logarithmic scale). Data points are from the billiard
	without ferrite in the frequency ranges 3--4~GHz (upper panel)
	and 9--10~GHz (lower panel). For clarity only every $5^{\rm
	th}$ data point is shown. The solid lines are best fits to the
	data. In the data shown in the lower panel the decay is
	dominated by noise for times larger than about 800~ns. These
	data are not taken into account in the fitting procedure.} 
	\label{fig:goe_fft}
\end{figure}

The measured scattering matrix elements are correlated for neighboring
frequencies $f$. The correlations result in a non-zero value of the
autocorrelation function defined in Eq.~(\ref{eqn:1}) and depicted in
Fig.~\ref{fig:goe_correl}. We show in Sec.~\ref{sec:expevidence} that
after a Fourier transformation the correlations between data points at
different times can be removed. This facilitates a statistically sound
analysis and is our motivation for using that transformation. Since
$S_{ab}(f)$ is measured at a discrete set of frequencies, the Fourier
coefficients $\tilde{S}_{ab}(k)$ are likewise obtained at discrete
time points $t_k = k / \Delta$, and the same is true of the
autocorrelation function $C_{a b}(\varepsilon)$ and its Fourier
transform $\tilde{C}_{a b}(k)$. Here, $\Delta =1$ GHz is
the length of the frequency interval and $k = 0, 1, \ldots, M - 1$,
see Section~\ref{fit}. We simplify the notation by using as
argument of the Fourier transforms the integer $k$. According to the
Wiener-Khinchin theorem we have $\tilde{C}_{a b}(k) =
|\tilde{S}_{ab}(k)|^2$. Figure~\ref{fig:goe_fft} shows two examples of
$\tilde{C}_{12}(k)$ at different values of $\Gamma / d$ for the
\Ti-invariant system. The solid lines in Fig.~\ref{fig:goe_fft} show a
fit of the random-matrix expression defined in Sec.~\ref{sec:theory}
to the data (the fit procedure is described in Sec.~\ref{fit}) and
correspond to the local-in-time mean values of the Fourier
coefficients. The data are seen to scatter about their time-dependent
mean. In Sec.~\ref{sec:expevidence} it is shown that the data points
divided by their local mean value at different times are indeed
uncorrelated and that the distribution of the rescaled Fourier
coefficients of the autocorrelation function is exponential. The decay
of the average function (solid line in Fig.~\ref{fig:goe_fft}) is
faster for $\Gamma/d=0.29$ (lower panel, frequency interval 9--10~GHz)
than for $\Gamma / d=0.02$ (upper panel, frequency interval 3--4~GHz).
This is due to stronger absorptive losses. In both cases the decay is
non-exponential (and the autocorrelation function is, therefore, not
Lorentzian). At $\Gamma / d=0.29$ and for times larger than about
1000~ns the decay is dominated by noise. Nevertheless, a decay over 5
orders of magnitude is experimentally well established.

\subsection{Elastic Enhancement Factor}

\begin{figure}[ht]
	\includegraphics[width=8cm]{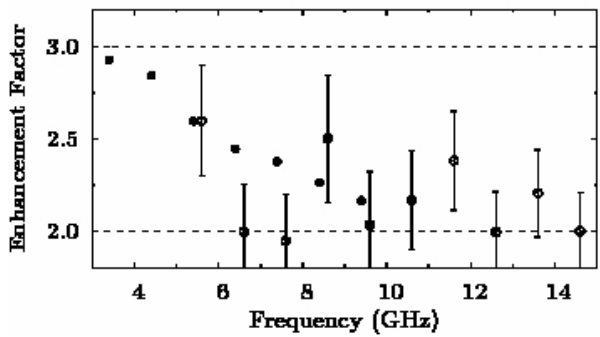} \caption{Elastic
	enhancement factors $\mathcal{W}$ for the \Ti-invariant
	billiard. The open circles are obtained with method (i), the
	filled circles with method (ii) described in the text. The
	error bars indicate uncertainties due to the finite range of
	data~\cite{Gibbs1965, Dallimore1966}. Above 10~GHz the analogy
	to a quantum billiard breaks down and Eq.~(\ref{eqn:Cabauto})
	needed for the analytic evaluation of $\mathcal{W}$ is no
	longer applicable.  The dashed horizontal lines indicate the
	limits of $\mathcal{W}$ for \Ti-invariant systems: Upper line
	for $\Gamma \ll d$, lower line for $\Gamma \gg d$.}
	\label{fig:enh_goe}
\end{figure}

\begin{figure}[ht]
	\includegraphics[width=8cm]{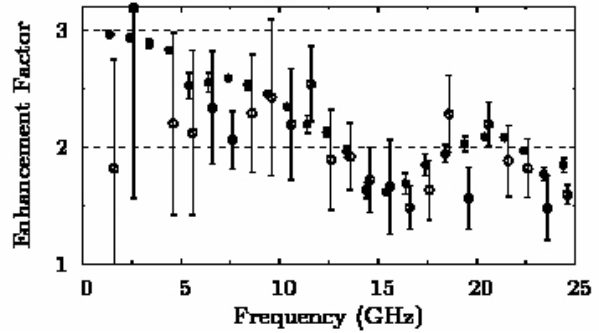} \caption{Elastic
	enhancement factors $\mathcal{W}$ for the billiard with
	partial TRIV. The open circles are obtained with method (i),
	the filled circles with method (ii) as described in the
	text. The error bars show the root-mean-square values for the
	6 realizations.  The dashed horizontal lines mark the limits
	of $\mathcal{W}$ for \Ti-invariant systems as in
	Fig.~\ref{fig:enh_goe}. Taken from Ref.~\cite{Schaefer2009}.}
	\label{fig:enh_190mT}
\end{figure}

In chaotic scattering, elastic processes are known to be
systematically enhanced over inelastic ones. The effect was first
found in nuclear physics~\cite{Satchler1963,EHK1969,Kretschmer1978}
but plays a role also in mesoscopic physics~\cite{Bergman}. The
enhancement depends on the degree of \T violation. The elastic
enhancement factor is defined as
\begin{eqnarray}
	\mathcal{W} &=& \sqrt{ \langle |S^{\rm fl}_{1 1}|^2\rangle \
        \langle |S^{\rm fl}_{2 2}|^2 \rangle } / \langle |S^{\rm fl}_{1
        2}|^2 \rangle \nonumber \\
        &=& \sqrt{C_{11}(0)\, C_{22}(0)}/C_{12}(0)\, ,
	\label{eqn:defW}
\end{eqnarray}
where the second equality results from Eq.~(\ref{eqn:1}). In the
limits of isolated resonances with many weakly coupled open channels
and of strongly overlapping resonances the values for $\mathcal{W}$
are~\cite{Savin2006}
\begin{equation}
	\mathcal{W} = \left\{
	\begin{array}{rl}
		1 + 2/\beta & \ {\rm for}\ \Gamma/d \ll 1 \\
		    2/\beta & \ {\rm for}\ \Gamma/d \gg 1\ .
	\end{array}
	\right.
	\label{eqn:Wlimits}
\end{equation}
Here, $\beta = 1$ for \Ti-invariant systems and $\beta = 2$ for
complete TRIV. The elastic enhancement factor $\mathcal{W}$ was
determined in two ways: (i) Using the first of Eqs.~(\ref{eqn:defW})
we calculated the averages over frequency directly from the
experimental values for $S_{ab}(f)$. This amounts to determine
$\mathcal{W}$ from a single experimental value for each of the
autocorrelation functions $C_{11}(0), C_{12}(0), C_{22}(0)$. (ii) In
the second of Eqs.~(\ref{eqn:defW}) we used the values of the
autocorrelation functions obtained by a best fit of the analytical
expression given in Eq.~(\ref{eqn:Cabauto}) below to the experimental
one. These are the solid lines in Fig.~\ref{fig:goe_fft}. The method
of fit (described in Sec.~\ref{fit}) uses the entire data set and is,
therefore, expected to give more reliable values for $\mathcal{W}$.
This is indeed borne out by the results shown in
Figs.~\ref{fig:enh_goe} and~\ref{fig:enh_190mT}. For the \Ti-invariant
case shown in Fig.~\ref{fig:enh_goe} the elastic enhancement factor
decreases from $\mathcal{W} \approx 3$ at low frequencies ($\Gamma \ll
d$) to $\mathcal{W} \approx 2$ at high frequencies ($\Gamma \approx
d$), in qualitative agreement with Eq.~(\ref{eqn:Wlimits}).

Results for the billiard with violated \T invariance are shown in
Fig.~\ref{fig:enh_190mT}. Although $\mathcal{W}$ was obtained from a
data set of 6 realizations, the values obtained with method (i) still
show large uncertainties while method (ii) yields reliable results.
Again $\mathcal{W}$ displays an overall decrease from $3$ to $2$ with
increasing $\Gamma / d$. However, at frequencies of about 6, 16 and
24~GHz dips are observed. Around 16 and 24~GHz the values of
$\mathcal{W}$ drop below 2. This is not possible for a \Ti-invariant
system. These features are similar to those of the cross-correlation
coefficient in Fig.~\ref{fig:crosscorrel}. Both measures indicate a
substantial violation of \Ti-invariance at about 6, 16 and 24~GHz.

\section{Theory}
\label{sec:theory}

As stated in the Introduction, it is the aim of the experiments
reported and analyzed in this paper to test random-matrix theory as
applied to chaotic scattering systems. According to the
Bohigas-Giannoni-Schmit conjecture~\cite{Boh84}, the spectral
fluctuation properties of chaotic \Ti-invariant quantum systems
coincide with those of the Gaussian orthogonal ensemble [GOE], those
of quantum systems with complete TRIV with those of the Gaussian
unitary ensemble [GUE] of random matrices. Systems with partial
violation of \T invariance are accordingly described by a crossover
from orthogonal to unitary symmetry. For such systems, analytical
expressions for central statistical measures of scattering processes
(i.e., the autocorrelation function~(\ref{eqn:1}) and the
cross-correlation coefficient~(\ref{eqn:cross})) have not been worked
out before. We fill that gap in Sec.~\ref{sec:autocorrel}.

\subsection{Crossover from Orthogonal to Unitary Symmetry: the
Autocorrelation function and the Cross-Correlation Coefficient}
\label{sec:autocorrel}

For chaotic scattering processes, the GOE $\to$ GUE crossover was
extensively investigated in Ref.~\cite{Pluhar1995}. There the
destruction of weak localization by an external magnetic field in the
transmission of electrons through a few-channel disordered
microstructure was determined. The connection between the conductance
$g$ and the $S$-matrix is given by the Landauer formula,
\begin{equation}
	g=\sum_{a=1}^{\Lambda /2}
	\sum_{b=\Lambda /2+1}^\Lambda \left \{ \vert
	S_{ab}\vert^2 + \vert S_{ba}\vert^2\right\}\, ,
	\label{eqn:cond}
\end{equation}
where $\Lambda$ counts the total number of open channels and an equal
number of incoming and outgoing channels $\frac{\Lambda}{2}$ is
assumed. Efetov's supersymmetry method~\cite{Efetov} was used to
calculate ensemble averages of squares of $S$-matrix elements $S_{a
b}$ for $a \neq b$. The $S$-matrix embodies the GOE $\to$ GUE
crossover in the manner described below. Here we describe the
extension of that approach to the calculation of the autocorrelation
function~(\ref{eqn:1}) and the cross-correlation
coefficient~(\ref{eqn:cross}). These observables were not considered
in Ref.~\cite{Pluhar1995}.

We write the unitary scattering matrix in the general
form~\cite{mahaux:69}
\begin{equation}
	S_{ab}(f) = \delta_{ab} - 2\pi i\sum_{\mu,\nu =1}^N W_{a\mu}
	\left[D^{-1}\right]_{\mu\nu}W_{b\nu}
	\label{eqn:Sab}
\end{equation}
where the inverse propagator $D$ is given by 
\begin{equation}
	D_{\mu\nu}=f\,\delta_{\mu\nu}-H_{\mu\nu} +
	i\pi\sum_{c=1}^\Lambda W_{c\mu}W_{c\nu}\, . \label{eqn:Dmn}
\end{equation}
The matrix elements $W_{a \mu}$ and $W_{b \mu}$ describe the coupling
of antennas $a$ and $b$ with the resonator mode
$\mu$~\cite{Schaefer2007}. The sum over $c$ in Eq.~(\ref{eqn:Dmn})
extends over the two antennas but includes also a number $(\Lambda -
2)$ of fictitious channels. The latter describe Ohmic absorption in
the cavity~\cite{Schaefer2003}. The matrix $H_{\mu \nu}$ is the
Hamiltonian of the closed billiard. It has dimension $N$ and the limit
$N \to \infty$ is eventually taken.

The coupling matrix elements $W_{c \mu}$ are chosen real, $W_{c \mu} =
W^*_{c \mu}$ for all $c, \mu$, and a violation of \T invariance by an
external magnetic field is taken into account only in $H_{\mu \nu}$.
We also assume that the $W_{c\mu}$ are independent of frequency
$f$. That assumption holds within frequency intervals of $1$ GHz
width. The experiment was designed such that direct power transmission
between the antennas is excluded so that the average $S$-matrix is
diagonal. We have verified that fact experimentally. A diagonal
average $S$-matrix is implied by the relation
\begin{equation}
	\sum_{\mu = 1}^N W_{a \mu} W_{b \mu}=N v_a^2 \delta_{a b}\, .
	\label{eqn:sumr}
\end{equation}
The parameter $v^2_a$ measures the average strength of the coupling of
the resonances to channel $a$. The Hamiltonian $H_{\mu \nu}$ is a
member of a random-matrix ensemble describing partial violation of
\T invariance. In random-matrix theory, the GOE $\to$ GUE
crossover is written as~\cite{Pandey}
\begin{equation}
	H_{\mu\nu}=H_{\mu\nu}^{\rm (S)}+i\frac{\pi\xi}{\sqrt{N}}
	H_{\mu\nu}^{\rm (A)}\, .
\label{eqn:hamiltonian}
\end{equation}
The real and symmetric matrix $H^{\rm (S)}$ is a member of the GOE,
and the elements of the real and antisymmetric matrix $H^{\rm (A)}$
are uncorrelated Gaussian-distributed random variables. Thus,
\begin{eqnarray}
        \langle H_{\mu\nu}^{\rm (S)} \rangle
		&=& \langle H_{\mu\nu}^{\rm (A)} \rangle = 0 \ ,
        \nonumber \\
        \langle H_{\mu\nu}^{\rm (S)} H_{\mu^\prime\nu^\prime}^{\rm (S)}
        \rangle &=&\frac{\lambda^2}{N}
	\left(\delta_{\mu\mu^\prime}\delta_{\nu\nu^\prime}
		+\delta_{\mu\nu^\prime}\delta_{\nu\mu^\prime}\right) \ ,
        \nonumber\\
	\langle H_{\mu\nu}^{\rm (A)}H_{\mu^\prime\nu^\prime}^{\rm (A)}
        \rangle &=&\frac{\lambda^2}{N}
	\left(\delta_{\mu\mu^\prime}\delta_{\nu\nu^\prime}
		-\delta_{\mu\nu^\prime}\delta_{\nu\mu^\prime}\right) \ .
	\label{eqn:var}
\end{eqnarray}
Here $\lambda$ has the dimension energy and for the GOE denotes half
the radius of Wigner's semicircle. The parameter $\xi$ measures the
strength of \Ti-invariance violation. For $\pi \xi / \sqrt{N} = 1$
the matrix $H$ is a member of the GUE. However, on the local level
(energy intervals measured in units of the mean level spacing $d$ of
the GOE) the transition from GOE to GUE already takes place when the
typical matrix element of the TRIV term becomes comparable to $d = \pi
\lambda / N$, i.e. when
\begin{equation}
	\frac{\pi \xi}{\sqrt{N}} \frac{\lambda}{\sqrt{N}} \simeq
	\frac{\pi}{\sqrt{N}} \frac{\lambda}{\sqrt{N}} \, \
\end{equation}
or when $\xi \simeq 1$. The $S$-matrix~(\ref{eqn:Sab}) is symmetric
only for $\xi = 0$, and reciprocity does not hold for $\xi \neq 0$.

Starting with Eq.~(\ref{eqn:Sflab}) we have used angular brackets to
denote the running average over (parts of) the experimental spectra.
The averages were actually taken over 1~GHz frequency intervals. Now
we consider an ensemble of $S$-matrix elements of the form of
Eq.~(\ref{eqn:Sab}) obtained by inserting many realizations of the
random Hamiltonian $H_{\mu\nu}$ in Eq.~(\ref{eqn:hamiltonian}). We
denote averages over that ensemble also by angular brackets. This
is legitimate because ergodicity guarantees the equality of ensemble
average and of the running average over a single realization of the
ensemble, see Sec. II C3 of \cite{HWM2009}.

The autocorrelation function for the $S$-matrix defined in
Eq.~(\ref{eqn:Sab}) is known for the case of \T invariance ($\xi =
0$)~\cite{Ver85}, for the case of complete TRIV ($\xi = \sqrt{N} /
\pi$)~\cite{Fyodorov2005} and had to be calculated for the case of
partial TRIV ($0 < \xi < \sqrt{N} / \pi$). For the first case it reads
\begin{eqnarray}
	C^{\rm GOE}_{ab}(\epsilon) &=& \frac{1}{8} \int_0^\infty{\rm
	d}\mu_1\int_0^\infty{\rm d}\mu_2\, \int_0^1 {\rm d}\mu \,
	\mathcal{J}(\mu, \mu_1, \mu_2) \nonumber \\ &\times&
	\exp\left(-i\frac{\pi \epsilon}{d} (\mu_1 + \mu_2 + 2 \mu)
	\right) \nonumber \\ &\times & \prod_c \frac{1-T_c\, \mu}
	{\sqrt{(1+T_c\mu_1) (1+T_c\, \mu_2)} } \nonumber \\ &\times&
	J_{ab}(\mu, \mu_1, \mu_2) \ .
\label{eqn:CabGOE}
\end{eqnarray}
The upper index of $C_{\rm ab}$ indicates that an average over the GOE
was taken. The integration measure is given by
\begin{eqnarray}
	\mathcal{J}(\mu,\mu_1,\mu_2) &=&\frac{\mu
	(1-\mu)\vert\mu_1-\mu_2\vert}{(\mu+\mu_1)^2 (\mu+\mu_2)^2}
	\nonumber \\ &\times&
	\frac{1}{\sqrt{(\mu_1(1+\mu_1)\mu_2(1+\mu_2)}} \ ,
	\label{lambdafkt}
\end{eqnarray} 
and we have
\begin{eqnarray}
	J_{ab}(\mu,\mu_1,\mu_2) &=& \delta_{ab} \
	\vert\left\langle{S_{aa}}\right\rangle\vert^2 \ T_a^2
	\nonumber \\ &\times& \bigg( \frac{\mu_1}{1+T_a\mu_1}
	\nonumber + \frac{\mu_2}{1+T_a\mu_2}+\frac{2\mu}{1-T_a\mu}
	\bigg)^2 \nonumber \\ &+& (1+\delta_{ab}) \ T_a T_b \ \bigg[
	\frac{\mu_1(1+\mu_1)} {(1+T_a\mu_1)(1+T_b\mu_1)} \nonumber \\
	&+& \frac{\mu_2(1+\mu_2)}{(1+T_a\mu_2)(1+T_b\mu_2)} \nonumber
	\\ &+& \frac{2\mu(1-\mu)}{(1-T_a\mu)(1-T_b\mu)} \bigg] \ .
	\label{jfkt}
\end{eqnarray}
Here and in Eqs.~(\ref{eqn:CabGUE}) and (\ref{eqn:Cab}) below the
input parameters are the mean level spacing $d$ and the transmission
coefficients $T_c$ in all channels $c$, defined as
\begin{equation}
	T_c = 1 - |\left\langle{S_{cc}}\right\rangle |^2 \ .
	\label{eqn:Tc}
\end{equation}
We observe that $0 \leq T_c \leq 1$. For all three cases the matrix
elements $W_{c\mu}$ occur in the final expression for the correlation
functions only via the transmission coefficients. Using the
analytical result for $\langle S_{a a} \rangle = (1 - \pi^2 v^2_c / d)
/ (1 + \pi^2 v^2_c /d)$ \cite{Ver85} and Eq.~(\ref{eqn:sumr}) one
finds that
\begin{equation}
	T_c = \frac{4 \pi^2 v^2_c / d}{(1 + \pi^2 v^2_c / d)^2} \ . 
\label{eqn:Tc1}
\end{equation}
The choice of the parameters $T_c$ is described in Section~\ref{fit}.
The threefold integrals in Eq.~(\ref{eqn:CabGOE}) and in
Eq.~(\ref{eqn:Cab}) below are numerically computed most conveniently
in terms of the integration variables introduced in
Ref.~\cite{Verbaarschot1986}.

For the second case, the $S$-matrix autocorrelation function was
worked out in Ref.~\cite{Fyodorov2005},
\begin{eqnarray}
	C^{\rm GUE}_{ab}(\epsilon) &=& \int_0^\infty{\rm
	d}\mu_1\int_0^1{\rm d}\mu \exp\left(-i\frac{2\pi \epsilon}{d}
	(\mu_1 + \mu)\right) \nonumber \\ &\times& \prod_c
	\frac{1-T_c\mu}{1+T_c\mu_1} \nonumber \\ &\times&
	\frac{T_a}{(1+T_a\mu_1)(1-T_a\mu)} \
	\frac{T_b}{(1+T_b\mu_1)(1-T_b\mu)} \nonumber \\
	&\times&\left(\delta_{ab}\vert\left\langle
	S_{aa}\right\rangle\vert^2 + \frac{1}{\mu_1+\mu}
	\left\{\mu_1-\mu + 1 \right.\right. \nonumber \\ && \left.-
	\mu_1 \mu \left(T_a + T_b - T_aT_b\right)\right\} \bigg) \ .
	\label{eqn:CabGUE}
\end{eqnarray}
The upper index of $C_{\rm ab}$ now indicates the average over the GUE.

In our experiments we deal with partial TRIV, i.e. with the third case
and the $S$-matrix autocorrelation function had to be calculated for
all values of the parameter $\xi$ introduced in
Eq.~(\ref{eqn:hamiltonian}). To this end we generalized the work of
Ref.~\cite{Pluhar1995}. We present here only the result and defer
details to the Appendix. The autocorrelation function is given in
terms of a threefold integral over integration variables $\lambda_0,
\lambda_1,$ and $\lambda_2$, see Eq.~(2) of Ref.~\cite{Gerland1996}.
However, the integrals are evaluated numerically more conveniently in
terms of the integration variables given in Sec. 5 of
Ref.~\cite{Verbaarschot1986}. For the transformation to these one
needs to distinguish in the integrations over $\lambda_1$ and
$\lambda_2$ the case where $\lambda_1 \ge \lambda_2$ and the case
where $\lambda_1 \le \lambda_2$. For instance, for the case
$\lambda_1\ge\lambda_2$ the transformation to integration variables
$\mu, \mu_1, \mu_2$ is given by
\begin{eqnarray}
	\lambda_0 &=& 1 - 2\mu\, , \nonumber \\ \lambda_1 &=& \sqrt{
	(1+\mu_1)\, (1+\mu_2)+\mu_1\mu_2 + \mathcal{U}}\, , \nonumber
	\\ \lambda_2 &=& \sqrt{
	(1+\mu_1)\,(1+\mu_2)+\mu_1\mu_2-\mathcal{U}}\, ,
	\label{eqn:Trafo}
\end{eqnarray}
where
\begin{equation}          
	\mathcal{U} = 2\sqrt{\mu_1 (1+\mu_1)\mu_2(1+\mu_2)}\, .
	\nonumber \\
\end{equation}
Then both, the $S$-matrix autocorrelation function and the
cross-correlation coefficient are obtained as special cases of a
function $F^\sigma_{a b}(\varepsilon)$. With the notations
\begin{equation}
	\mathfrak{t} = \pi^2\,\xi^2,\, 
\end{equation}
and
\begin{eqnarray}
	\mathcal{R} &=& 4(\mu + \mu_1)(\mu + \mu_2)\, ,
	\label{eqn:defs} \\ \mathcal{F} &=& 4\mu(1-\mu),\
	\mathcal{G}=\lambda_1^2-1,\ \mathcal{H}=\lambda_2^2-1,
	\nonumber\\ \varepsilon_\pm &=& 1 \pm
	\exp(-2\mathfrak{t}\mathcal{F}), \nonumber\\ \tilde{A}_a &=&
	\frac{(2-T_a) \lambda_2 + T_a \lambda_1}{4\, (1 + T_a
	\lambda_1) (1 + T_a \lambda_2)}, \nonumber \\ \tilde{B}_a &=&
	\frac{(2-T_a) \lambda_1 + T_a \lambda_2}{4\, (1 + T_a
	\lambda_1) (1 + T_a \lambda_2)}, \nonumber \\ \tilde{C}_a &=&
	\frac{1}{2}\frac{1}{1-T_a \mu}, \, C_1 =
	\frac{\mu\,(1-\mu)}{(1-T_a\mu)\,(1-T_b\mu)}, \nonumber \\ C_2
	&=& \frac{\mathcal{U}}{4} \left( \frac{1}{1 + T_a\mu_2}\,
	\frac{1}{1 + T_b\mu_1} \right. \nonumber \\ && \left. +
	\frac{1}{1+T_a\mu_1}\, \frac{1}{1 + T_b \mu_2} \right),
	\label{eqn:Notat}
\end{eqnarray}
the function $F^\sigma_{a b}(\varepsilon)$ reads
\begin{eqnarray}
	F^\sigma_{ab}(\epsilon) &=& \frac{1}{8} \int_0^\infty{\rm
	d}\mu_1\int_0^\infty{\rm d}\mu_2\, \int_0^1 {\rm d}\mu\,
	\frac{\mathcal{J}(\mu,\mu_1,\mu_2)}{\mathcal{F}} \nonumber \\
	& \times&\exp\left(-\frac{i \pi \epsilon}{d} (\mu_1 + \mu_2 +
	2 \mu)\right) \nonumber \\ &\times& \prod_c \frac{1-T_c\,
	\mu}{\sqrt{(1+T_c\mu_1) (1+T_c\,\mu_2)} } \nonumber\\
	&\times&\Big[\exp\left( -2\mathfrak{t}\mathcal{H}\right)\cdot
	\Big\{ J_{ab}(\mu,\mu_1,\mu_2)\cdot \left[ \mathcal{F}
	\varepsilon_+ \right. \nonumber\\ &+& (\lambda_2^2 -
	\lambda_1^2) \varepsilon_- + \left.4\mathfrak{t}\mathcal{R}
	(\lambda_2^2\varepsilon_- + \mathcal{F} (\varepsilon_+ - 1))
	\right] \nonumber \\ &+& \sigma\cdot 2(1 -
	\delta_{ab})T_aT_bK_{ab}\Big\} +
	\left(\lambda_1\leftrightarrow\lambda_2\right)\Big]\,.
\label{eqn:Cab}
\end{eqnarray}
Here, 
\begin{eqnarray}
	K_{ab} &=& \varepsilon_- \Big[ 2\mathfrak{t} \mathcal{R} C_1
	\mathcal{F} \label{eqn:Kab} \\ &+& 2\mathcal{F} \left\{
	(\tilde{A}_a \tilde{C}_b + \tilde{A}_b \tilde{C}_a)\mathcal{G}
	\lambda_2 + (\tilde{B}_a \tilde{C}_b + \tilde{B}_b
	\tilde{C}_a)\mathcal{H} \lambda_1 \right\} \nonumber \\ &+& 3
	C_1 \mathcal{F} - C_2 (\lambda_2^2 - \lambda_1^2) + C_2
	\mathfrak{t} \mathcal{R} (4\lambda_2^2 - 2\mathcal{F}) \Big]
	\nonumber \\ &+& \left( \varepsilon_+ -
	\frac{\varepsilon_-}{\mathfrak{t} \mathcal{F}} \right) \Big[ 3
	C_1 (\lambda_2^2 - \lambda_1^2) + \mathfrak{t} \mathcal{R} C_1
	(4 \lambda_2^2 - 2\mathcal{F}) \nonumber \\
	&+&2\mathcal{F}\left\{ (\tilde{A}_a \tilde{C}_b + \tilde{A}_b
	\tilde{C}_a)\mathcal{G}\lambda_2 - (\tilde{B}_a \tilde{C}_b +
	\tilde{B}_b \tilde{C}_a)\mathcal{H}\lambda_1 \right\}
	\nonumber \\ &+&(2\mathfrak{t}\mathcal{R}-1)C_2\mathcal{F}
	\Big]. \nonumber
\end{eqnarray}
The integration measure $\mathcal{J}(\mu,\mu_1,\mu_2)$ and the
function $J_{ab}(\mu,\mu_1,\mu_2)$ are given explicitly in
Eqs.~(\ref{lambdafkt}) and Eq.~(\ref{jfkt}). Setting $\sigma = +$
($\sigma = -$) in Eq.~(\ref{eqn:Cab}) yields the autocorrelation
function (the cross-correlation coefficient),
\begin{eqnarray}
	C_{ab}(\epsilon) &=& 
		F^+_{ab}(\epsilon)\label{eqn:Cabauto}\\
	C_{\rm cross}(\epsilon=0) &=& 
		F^-_{ab}(\epsilon=0)\, .\label{eqn:Cabcross}
\end{eqnarray}

\begin{figure}[ht]
	\includegraphics[width=8cm]{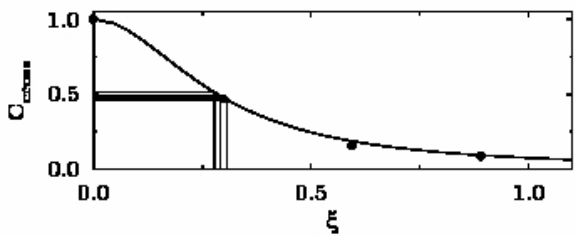}
	\caption{Dependence of the cross-correlation coefficient
	$C_{\rm cross}(0)$ on the parameter $\xi$ as predicted by a
	random-matrix model for partial violation of \T
	invariance. The analytic result (line) is compared with an RMT
	simulation (dots) for the same set of transmission
	coefficients. Also shown is how an experimental value of
	$C_{\rm cross}(0) = 0.49(3)$, c.f.\
	Fig.~\ref{fig:crosscorrel}, translates into
	$\xi=0.29(2)$. Based on Ref.~\cite{Schaefer2009}.}
	\label{fig:cross_vs_xi}
\end{figure}

\begin{figure}[ht]
	\includegraphics[width=8cm]{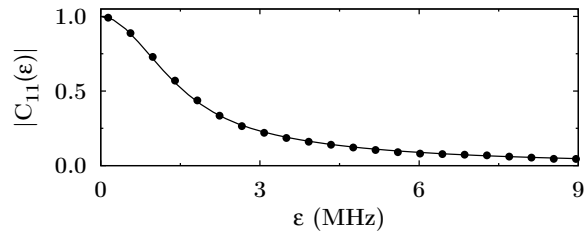}
	\caption{Comparison of the analytic result for the
	autocorrelation function $C_{1 1}$ versus $\varepsilon$ for
	transmission coefficients $T_1 = 0.407$, $T_2 = 0.346$,
	$\tau_{\rm abs} = 2.41$ and TRIV parameter $\xi =0.293$ (solid
	line) with RMT simulations (dots). We show only the result for
	$C_{1 1}$ as that for $C_{1 2}$ is barely distinguishable. The
	curve is normalized such that it equals unity for
	$\varepsilon=0$.}  \label{fig:correl_vs_rmt_f}
\end{figure}

\begin{figure}[ht]
	\includegraphics[width=8cm]{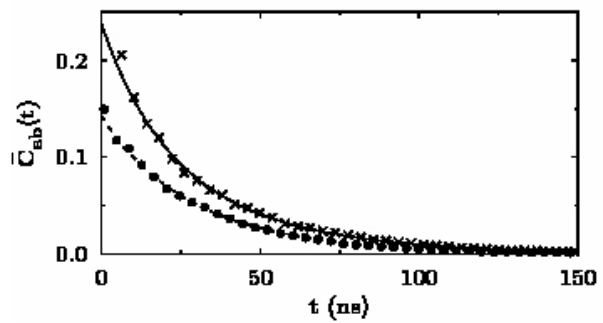}
	\caption{Comparison of the analytic results for the Fourier
	transform of the autocorrelation function versus time $t$ for
	transmission coefficients $T_1 = 0.407$, $T_2 = 0.346$,
	$\tau_{\rm abs} = 2.41$ and TRIV parameter $\xi = 0.293$ to
	RMT simulations. We show the results for $\tilde{C}_{a b}$ for
	$a=b=1$ (dashed line and filled points) and $a=1,\ b=2$ (solid
	line and crosses, respectively).}  \label{fig:correl_vs_rmt_t}
\end{figure}

We observe that for $\xi =0$, i.e. $\mathfrak{t}=0$, the function
$K_{ab}$ defined in Eq.~(\ref{eqn:Kab}) vanishes and
$F^+_{ab}(\epsilon)$ in Eq.~(\ref{eqn:Cab}) turns into the
autocorrelation function of the GOE given in Eq.~(\ref{eqn:CabGOE}).
We checked our analytical results by comparison with RMT simulations.
In Fig.~\ref{fig:cross_vs_xi} we show the cross-correlation
coefficient versus $\xi$ as obtained analytically and by RMT
simulation for a typical set of transmission coefficients. We also
indicate how the analytical result is used to determine the value of
$\xi$ from a measured value of the cross-correlation coefficient. To
test the validity of Eq.~(\ref{eqn:Cab}) we compare in
Figs.~\ref{fig:correl_vs_rmt_f} and \ref{fig:correl_vs_rmt_t} analytic
results for the autocorrelation functions with numerical simulations,
both in the frequency and in the time domains. The parameter
$\tau_{\rm abs}$ measures absorption and is defined in
Sec.~\ref{fitpar} below. In all cases, the agreement is very good.

The theoretical expressions given in Eqs.~(\ref{eqn:CabGOE}),
(\ref{eqn:CabGUE}) and (\ref{eqn:Cab}) are obtained as averages over
the ensemble of Hamiltonian matrices defined in
Eq.~(\ref{eqn:hamiltonian}) in the limit $N \to \infty$ and directly
yield the autocorrelation function. In contrast,
Fig.~\ref{fig:goe_correl} shows autocorrelation functions obtained by
averaging the data in a frequency interval of $1$~GHz width. For
conceptual clarity we distinguish both cases by referring to the
theoretical and to the experimental autocorrelation functions,
respectively.

\subsection{Parameters}
\label{fitpar}

\begin{figure}[ht]
	\includegraphics[width=8cm]{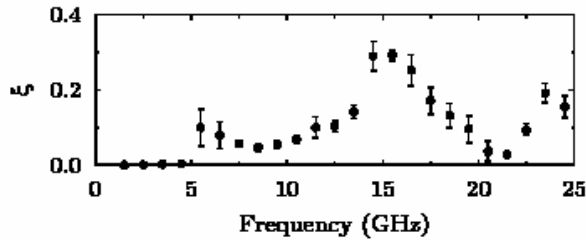} \caption{Values of the
	TRIV parameter $\xi$ for the billiard with the ferrite
	magnetized with $B=190~{\rm mT}$. The error bars indicate the
	variability of the results within the 6 realizations.}
	\label{fig:xi}
\end{figure}

\begin{figure}[ht]
	\includegraphics[width=8cm]{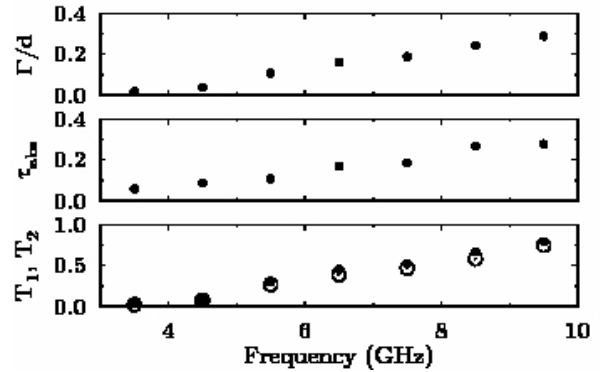}
	\caption{$\Gamma/d$ (top panel), $\tau_{\rm abs}$ (middle
	panel) and the transmission coefficients $T_1$ (filled
	circles) and $T_2$ (open circles) (bottom panel) versus
	frequency for the billiard with \T invariance. The errors are
	typically of the size of the symbols.} 
	\label{fig:Tcoeff_goe}
\end{figure}

\begin{figure}[ht]
	\includegraphics[width=8cm]{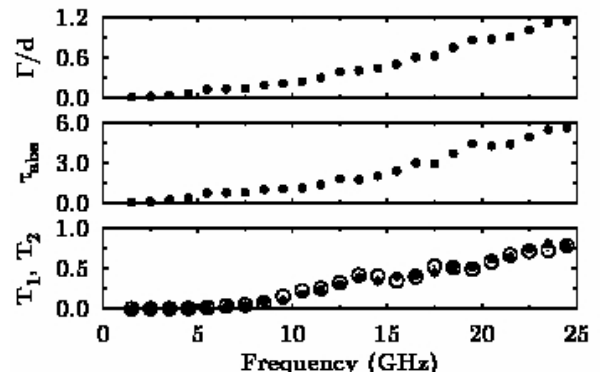} \caption{The
	same as in Fig.~\ref{fig:Tcoeff_goe} but for the billiard with
	the ferrite, obtained as averages over six realizations. The
	scatter of the values for different realizations about the
	mean value is typically of the order of the symbol size.}
\label{fig:Tcoeff_190mT}
\end{figure}

The parameters in Eqs.~(\ref{eqn:CabGOE}), (\ref{eqn:CabGUE}) and
(\ref{eqn:Cab}) are the average level spacing $d$, the transmission
coefficients $T_c$ for all channels $c$, and the parameter $\xi$ for
TRIV. We have calculated $d$ from the Weyl formula~\cite{Bal76}. A
starting value for the parameter $\xi$ was determined from the
experimental cross-correlation coefficients shown in
Fig.~\ref{fig:crosscorrel} as described in the caption of
Fig.~\ref{fig:cross_vs_xi}. Here we use that the
cross-correlation coefficient depends only weakly on the transmission
coefficients in the frequency range 1--25~GHz. Results are shown in
Fig.~\ref{fig:xi}.
The largest value of $\xi$ is $\xi\simeq 0.3$. In determining
$T_c$ by fitting the theoretical expressions for the autocorrelation
function to the data, we also use $\xi$ as fit parameter, with
starting value as just described. For the channels $c = 1$ and $c =
2$, i.e.\ for the antennas, we also have determined starting values
from the definition Eq.~(\ref{eqn:Tc}) and from the measured values of
$S_{11}(f)$ and $S_{22}(f)$. The fits discussed in
Sec.~\ref{sec:FTdistrib} yielded $T_1$, $T_2$ within 5~\% of these
starting values. The remaining transmission coefficients describe
Ohmic absorption in the walls of the resonator and the ferrite. If
$T_c \ll 1$ for all absorbing channels, the products over these
channels appearing in Eqs.~(\ref{eqn:CabGOE}), (\ref{eqn:CabGUE})
and~(\ref{eqn:Cab}) simplify so that each of the three theoretical
autocorrelation functions depends only on the sum $\tau_{\rm abs}$ of
the transmission coefficients for the absorbing channels. Accordingly,
in addition to $T_1$, $T_2$ and $\xi$ the parameter $\tau_{\rm abs}$
was used as fitting parameter. To estimate the correlation width
$\Gamma$ of the resonances, we have used the Weisskopf formula~\cite{Bla52}
\begin{equation}
	2\,\pi\, \frac{\Gamma}{d} = \sum_c T_c = T_1 + T_2 +
        \tau_{\rm abs}\ .
	\label{eqn:GammaD}
\end{equation}
and the fitted values for $T_1$, $T_2$, and $\tau_{\rm abs}$. 
Using numerical simulations and with help of the analytic result
Eq.~(\ref{eqn:CabGOE}) for the autocorrelation function we checked that
this formula indeed yields a very good estimate for the correlation
length even in the regime of weakly overlapping resonances and for a
few open channels. The results of our fits (each done in a frequency
interval of $1$ GHz length) are displayed in
Figs.~\ref{fig:Tcoeff_goe} and \ref{fig:Tcoeff_190mT}. We show $\Gamma
/ d$ as obtained from Eq.~(\ref{eqn:GammaD}) (top panel), $\tau_{\rm
abs}$ (middle panel) and $T_1$ and $T_2$ (bottom panel) versus
frequency for the case without and with TRIV, respectively. The
transmission coefficients and $\tau_{\rm abs}$ generally increase with
frequency. We note that without the ferrite $\Gamma / d$ never exceeds
the value $0.3$ (regime of weakly overlapping resonances) since the
excitation frequency must be chosen below $f_{\rm max}=10.3$~GHz,
whereas in the system with ferrite we have $f_{\rm max}=30$~GHz and
thus $\Gamma / d$ attains values as large as $1.2$.

\subsection{Distribution of $S$-Matrix Elements}
\label{sec:Sdistrib}

The distribution of $S$-matrix elements is not known completely in
analytical form, neither for the \Ti-invariant system nor for TRIV.
The most complete information is available in the regime of strongly
overlapping resonances ($\Gamma /d \gg 1$). In a basis in channel
space where $\langle S_{a b} \rangle$ is diagonal, the inelastic
$S$-matrix elements $S_{a b} = S^{\rm fl}_{a b}$ with $a \neq b$ have
a bivariate Gaussian distribution~\cite{Aga75}. Thus the phase of
$S_{a b}$ is uniformly distributed in the interval $\{- \pi, \pi\}$.
The distribution function $P$ of the modulus $r = |S_{a b}|$ depends
only on the ratio $z = r / (\langle |S_{a b}|^2 \rangle)^{1/2}$ and is
given by
\begin{equation}
	P(z) = \frac{\pi}{2}z \exp\left[-\frac{\pi}{4} z^2\right] \ .
	\label{vertsab}
\end{equation}
The diagonal elements $S_{a a}$ have a bivariate Gaussian distribution
only for $\Gamma \gg d$ and only if $| \langle S_{a a} \rangle| \ll 1$.
Otherwise, unitarity constraints cause the distribution to differ from
the Gaussian form~\cite{Dav88,Dav89,Dietz2009a}.

In Ref.~\cite{Fyodorov2005} an analytic expression for the
distribution of the elastic elements $S_{a a}$ of the scattering
matrix of a generic chaotic system without or with partially violated
\T invariance was derived. It applies for cases with many open
channels. With the notation
\begin{equation}
	S_{a a} = \sqrt{r_a}e^{i\theta_a},\, x_a =
	\frac{r_a+1}{r_a-1},\, g_a = \frac{2}{T_a} - 1,
\end{equation}
the distribution $P(x_a,\theta_a)$ of $S_{a a}$ is given by 
\begin{eqnarray}
	P(x_a,\theta_a) &=& \frac{1}{2\,\pi} \frac{d}{dy}(y^2-1)
	\nonumber \\ &\times& \left. \frac{d\mathfrak{f}^a(y)}{dy}
	\right\vert_{y = x_ag_a + \sqrt{x_a^2-1} \sqrt{g_a^2-1}
	\cos\theta_a}\,.  \label{eqn:Scc_distrib}
\end{eqnarray}
With the help of the transformation Eq.~(\ref{eqn:Trafo}), the
definitions~(\ref{eqn:Notat}) and with $w = (y - 1) / 2$ the analytic
expression for the function $\mathfrak{f}^a(y)$ given in
Ref.~\cite{Fyodorov2005} may be cast into the form
\begin{widetext}
\begin{eqnarray}
	\mathfrak{f}^a(y) &=& \frac{1}{4} \int_0^{\rm w_c}{\rm
	d}\mu_2\int_{w_c}^\infty{\rm d}\mu_1\, \int_0^1 {\rm d}\mu\,
	\frac{\mathcal{J}(\mu,\mu_1,\mu_2)}{\mathcal{F}}
	\label{eqn:Fx} \\ & \times&\exp\left(-\frac{\tau_{\rm
	abs}}{2}(\mu_1 + \mu_2 + 2 \mu)\right)
	\frac{w+\mu}{\sqrt{(w-\mu_1)(\mu_2-w)}} \prod_{c\ne
	a}\frac{1-T_c\, \mu}{\sqrt{(1+T_c\mu_1) (1+T_c\,\mu_2)} }
	\nonumber\\ &\times& \Big[ \exp\left( -2 \mathfrak{t}
	\mathcal{H} \right) \cdot \left[ \mathcal{F} \varepsilon_+ +
	(\lambda_2^2 - \lambda_1^2) \varepsilon_-
	+4\mathfrak{t}\mathcal{R} (\lambda_2^2 \varepsilon_- +
	\mathcal{F} (\varepsilon_+ - 1))\right] +
	\left(\lambda_1\leftrightarrow\lambda_2\right)\Big]. \nonumber
\end{eqnarray}
For \Ti-invariant systems the threefold integral can be simplified,
and $P(x_a,\theta_a)$ takes the form
\begin{eqnarray}
	P(x_a,\theta_a) &=& \frac{1}{4\pi} \frac{d}{dy} \left(1 + y
	\right) \Big[\tau_{\rm abs} \left(K_1(w) J_2(w) + K_2(w)
	J_1(w) \right) \nonumber\\ &+& \left.\sum_{c=1}^\Lambda t_c^a
	\left(L_1^c(w) H_2^c(w) + L_2^c(w) H_1^c (w) \right)
	\Big]\right\vert_{y= x_a g_a + \sqrt{x_a^2-1} \sqrt{g_a^2-1}
	\cos\theta_a}\, .
\label{result}
\end{eqnarray}
Here, $t_c^a=1$ for $c=a$ and $t_c^a = T_c$ otherwise, $\Lambda$ is
the number of open channels and
\begin{eqnarray}
	J_1(w) &=& \int_w^\infty {\rm d}y \frac{e^{-\tau_{\rm abs}\,
	y/2}}{\sqrt{y|y-w|}} \prod_{d=1}^\Lambda \frac{1}{\sqrt{1
	+t_d^a\, y}}, \nonumber\\ H_1^{c}(w) &=& \int_w^\infty {\rm
	d}y \frac{e^{-\tau_{\rm abs}\, y/2}}{\sqrt{y|y-w|}}
	\prod_{d=1}^\Lambda \frac{1}{\sqrt{1 +t_d^a\, y}} \frac{1}{1 +
	t_c^a\, y}, \nonumber\\ K_1(w) &=& \int_w^\infty {\rm d}y\,
	e^{-\tau_{\rm abs}\, y/2}
	\frac{\sqrt{y|y-w|}}{\prod_{d=1}^\Lambda \sqrt{1 + t_d^a\, y}}
	\left[\frac{e^{-\tau_{\rm abs}}}{y + 1} \prod_{d=1}^\Lambda
	\left(1 - t_d^a\right) - \frac{1}{y} \right.\nonumber\\ &+&
	\left. \sum_{b=1}^\Lambda \frac{{t_b^a}^2}{1+t_b^ay} \int_0^1
	{\rm d}\mu_0\, e^{-\tau_{\rm abs} \mu_0} \prod_{d\ne
	b}^\Lambda \left( 1 - t_d^a\mu_0 \right) \right], \nonumber \\
	L_1^{c}(w) &=& \int_w^\infty {\rm d}y\, e^{-\tau_{\rm abs}\,
	y/2} \frac{\sqrt{y|y-w|}}{\prod_{d=1}^\Lambda \sqrt{1 +
	t_d^a\, y}} \left[ \frac{e^{-\tau_{\rm abs}}}{y + 1}
	\prod_{d\ne c}^\Lambda \left(1 - t_d^a\right) - \frac{1}{y}
	\right. \nonumber\\ &+& \left. \sum_{b\ne c}
	\frac{{t_b^a}^2}{1 + t_b^ay} \int_0^1 {\rm d}\mu_0\,
	e^{-\tau_{\rm abs} \mu_0} \prod_{d\ne b,c}^\Lambda \left(1 -
	t_d^a \mu_0 \right) \right]\, .  \label{S11integral}
\end{eqnarray}
\end{widetext}
The corresponding functions with index 2 are given by the same
expression except that the integration limits $w, \infty$ have to be
replaced by $0, w$.

\begin{figure}[ht]
	\includegraphics[width=8cm]{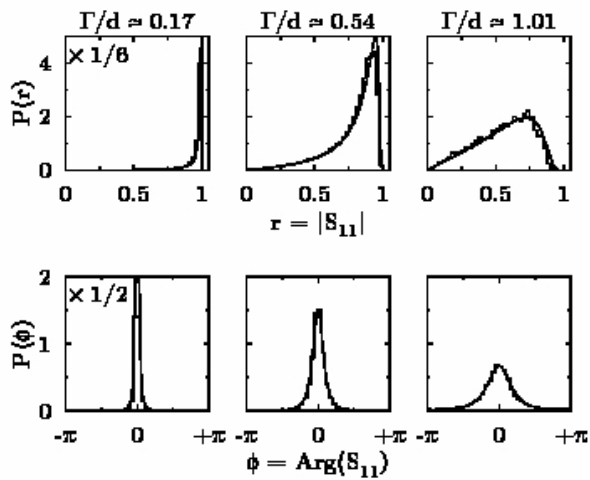}
	\caption{Distribution of $S_{11}$-matrix elements according
	to modulus (upper panels) and phase (lower panels). The
	histograms give the probability distribution functions in the
	frequency ranges (from left to right) 9--10~GHz, 17--18~GHz
	and 23--24~GHz. The data were measured with the billiard used
	for TRIV but without ferrite and for a total of 8
	realizations, i.e.\ each graph is constructed from $80\,000$
	data points. All of these were used in the histograms, and
	correlations between $S$-matrix elements at neighboring
	energies were thus neglected. Nevertheless, the analytical
	result Eq.~(\ref{eqn:Scc_distrib}) (solid lines) agrees with
	the data.}
	\label{fig:S11_distrib}
\end{figure}

These analytic results were previously tested experimentally for the
case of a single open channel plus absorption in
Refs.~\cite{Mendez2003,Kuhl2005,Hul2005,Lawniczak2008,HarmInv}. In
Fig.~\ref{fig:S11_distrib} we compare for several frequency intervals
the experimental distributions of the elastic $S$-matrix element
$S_{11}$ with the theoretical predictions for the case of two open
channels with absorption. The data were taken with the billiard used
for the experiments with partial TRIV but without the ferrite. This
was done because that billiard has a smaller height so that the range
where only a single vertical mode is excited, extends up to $30$~GHz.
Higher values of the frequency result in larger absorption and in
larger values of $\Gamma / d$. It is here that the theoretical
result~(\ref{result}) is expected to apply. The value of $\Gamma / d$
was determined from the Weisskopf formula given in 
Eq.~(\ref{eqn:GammaD}) and from
the values of $T_1,\,T_2,\,\tau_{\rm abs}$ obtained from a fit of the
Fourier transform of the $S$-matrix autocorrelation function as
described in Sec.~\ref{sec:FTdistrib}. The very good agreement
corroborates the precision of the fitting procedure and of the GOF
test discussed below. Note that the distributions are very far from a
bivariate Gaussian distribution.

\begin{figure}[ht]
	\includegraphics[width=8cm]{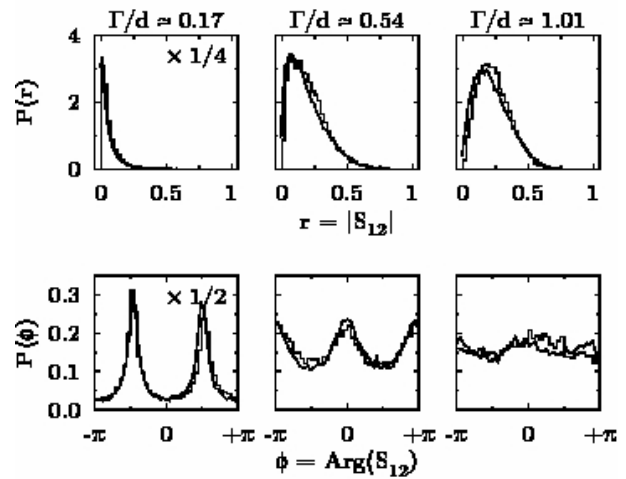}
	\caption{Distribution of $S_{12}$-matrix elements according
	to modulus (upper panels) and phase (lower panels) for data
	points as described in the caption of
	Fig.~\ref{fig:S11_distrib}. An RMT simulation (solid lines)
	shows acceptable agreement with the data.}
	\label{fig:S12_distrib}
\end{figure}

An analytic expression for the distribution of the off-diagonal
elements of the $S$-matrix exists in the Ericson regime $\Gamma \gg
d$, however not in the range of $\Gamma /d$ achieved in the
experiments. In Fig.~\ref{fig:S12_distrib} we, therefore, compare
experimental distributions to RMT simulations. We note again the good
agreement. In the frequency range 23--24~GHz (where $\Gamma/d \approx
1.01$) the distribution of $|S_{1 2}|$ is well described by
Eq.~(\ref{vertsab}), and the distribution of the phases is nearly
uniform. Thus, in this frequency range the distribution of the
non-diagonal $S$-matrix elements is already close to that expected in
the Ericson regime while that of the diagonal elements is still far
from Gaussian.

\section{Data Fits and Distribution of Fourier Coefficients}
\label{sec:FTdistrib}

In the present Section we test predictions of random-matrix theory
with the experimental data. We proceed as follows. Using the results
of Sec.~\ref{fourtrafo}, we fit the parameters of the Fourier
transforms of the theoretical expressions for the autocorrelation
functions (Eqs.~(\ref{eqn:CabGOE}) and~(\ref{eqn:Cab})) to the data.
We show that within the accuracy of the data and after rescaling, the
distribution of the Fourier-transformed $S$-matrix elements is
Gaussian. This property is used to develop a goodness-of-fit (GOF)
test that quantitatively tests the quality of RMT predictions. As a
second test of RMT we compare predicted values of the elastic
enhancement factors with the data.

In Sec.~\ref{sec:Sdistrib} we have shown that the real and imaginary
parts of the $S$-matrix elements in general do not have a Gaussian
distribution. How can this fact be reconciled with the statement just
made that the Fourier-transformed $S$-matrix elements do have such a
distribution? The Fourier transform is a linear transformation, after
all. We answer that question as we proceed.

\subsection{Fits}
\label{fit}

We focus attention on the fluctuating part $S^{\rm fl}(f)$ of the
$S$-matrix elements (see Eq.~(\ref{eqn:Sflab})) and omit the indices
$a$ and $b$ for brevity. By definition we have $\langle S^{\rm fl}(f)
\rangle = 0$. Data are taken at frequency increments $\Delta\geq
100~{\rm kHz}$. The mean level spacing $d$, the transmission
coefficients $T_1$ and $T_2$, and the absorption coefficient $\tau_{\rm
abs}$ are typically constant in frequency intervals of $1$ GHz width.
In every such interval we have $M \simeq 10^4$ measured values of $S^{\rm
fl}(f)$ for all combinations of channel indices $a, b$. We write $f_j =
f_0 + j \Delta$ where $f_0$ is the frequency at the lower end of the
interval and $j = 0, 1, 2, \ldots, (M-1)$. We use discrete Fourier
transformation and define
\begin{equation}
	\tilde{S}_k = \sum_{j = 0}^{M-1} e^{-2 \pi
	i\,k\,j/M}S^{\rm fl}(f_j), \ k = 0, \ldots, M-1\, ,
	\label{eqn:ftdiscret}
\end{equation}
so that
\begin{equation}
	S^{\rm fl}(f_j) = \frac{1}{M} \sum_{k=0}^{M-1} e^{2
        \pi i\,k\,j/M} \tilde{S}_k,\ j=0, \ldots, M-1\, .
	\label{eqn:ftinverse}
\end{equation}
We fit the parameters in the theoretical
expressions~(\ref{eqn:CabGOE}) and (\ref{eqn:Cab}) to the distribution
of the squares $x_k = |\tilde{S}_k|^2$ of these Fourier coefficients.
The Wiener-Khinchin theorem states that the latter are equal to the
Fourier coefficients $\tilde{C}(k)$ of the experimental
autocorrelation functions $C(\varepsilon)$. We accordingly calculate
the Fourier transforms $\tilde{C}(k)$ of the theoretical
autocorrelation functions in Eqs.~(\ref{eqn:CabGOE}) and
(\ref{eqn:Cab}) at the same discrete values of $k$ as occur in the
discrete Fourier transformation in Eq.~(\ref{eqn:ftdiscret}). The
parameters are the transmission coefficients $T_1$, $T_2$, $\tau_{\rm
abs}$ and $\xi$. For the transmission coefficients $T_1$ and $T_2$ we
used Eq.~(\ref{eqn:Tc}) with experimental values for $\langle S_{a a}
\rangle $ and $a = 1, 2$ as starting points but allowed the values of
$T_1$ and $T_2$ to vary. The best-fit values differed by no more than
$5$~\% from the starting values. For $\xi$ a starting value was
obtained from a comparison of the experimental and the analytic
cross-correlation coefficients as outlined in Sec.~\ref{fitpar}. For
the fit parameter $\tau_{\rm abs}$ no starting values could be
computed from the measured data. The fit yields the solid lines shown
in Fig.~\ref{fig:goe_fft} and defines $\langle x_k \rangle$. We
observe that $\langle x_k \rangle$ decreases by several orders of
magnitude over the available range of $k$ values. We also observe that
for $k \neq k'$ the Fourier coefficients $\tilde{S}_k$ and
$\tilde{S}_{k'}$ are uncorrelated. This follows from the fact that the
autocorrelation functions depend only on the difference $\varepsilon$
of the two frequency arguments and is shown below in Eq.~(\ref{eqn:7}).

As stated in the Introduction, the use of generic expressions derived
from RMT is justified only for energy spacings bounded from above by
the period of the shortest periodic orbit in the classical microwave
billiard. Therefore, Fourier coefficients are generic only for times
larger than the repetition time of the shortest periodic orbit We have
estimated that time and found it to be smaller than the time-index $k
= 5$ of $\tilde{S}_k$ in Eq.~(\ref{eqn:ftdiscret}). We note, however,
that the average $S$-matrix elements (which correspond to $k = 0$) are
not generic. This is mirrored by the fact that these are used as input
parameters in our analysis.

\subsection{Gaussian Distribution}
\label{sec:expevidence}

\begin{figure}[ht]
	\includegraphics[width=8cm]{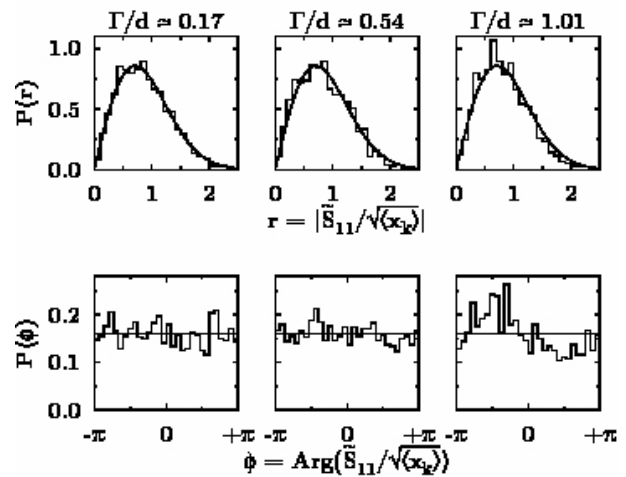}
	\caption{Distribution $P$ of the rescaled Fourier coefficients
	$\tilde{S} / \sqrt{\langle x_k \rangle }$ for the elastic case,
	$\{a, b\} = \{1, 1\}$. Upper panels: The data (histograms) for the
	distribution agree well with the solid lines given by
	Eq.~(\ref{vertsab}). Lower panels: The phases are uniformly
	distributed in the interval $\{ - \pi, + \pi\}$. Data source is as
	in Fig.~\ref{fig:S11_distrib} with Fourier coefficients taken from
	the first 200~ns, i.e.\ $1\,240$ data points contribute to each
	histogram.}
	\label{fig:S11fft_distrib}
\end{figure}

\begin{figure}[ht]
	\includegraphics[width=8cm]{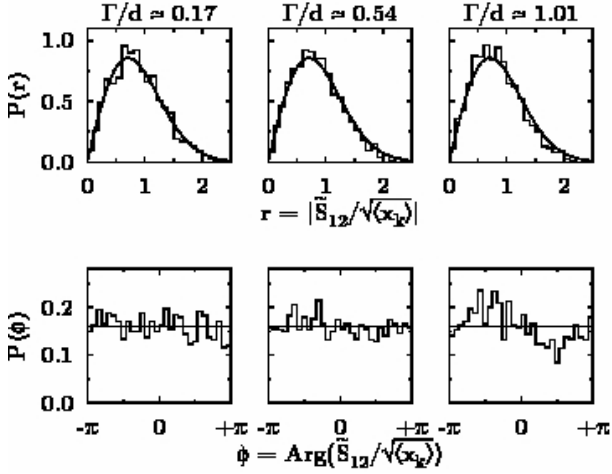} 
	\caption{Same as in Fig.~\ref{fig:S11fft_distrib} but for the
	inelastic case $\{a, b\} = \{1, 2\}$. Data source is as in
	Fig.~\ref{fig:S12_distrib} with Fourier coefficients taken
	from the first 200~ns.}
	\label{fig:S12fft_distrib}
\end{figure}

\begin{figure}[ht]
	\includegraphics[width=8cm]{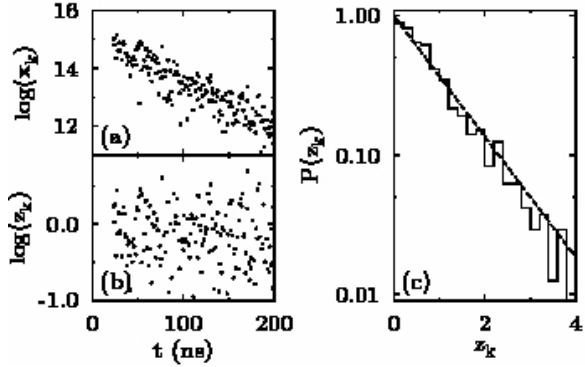} 
	\caption{Distribution of the Fourier coefficients $x_k =
	\vert \tilde{S}_{12} \vert^2$ in the interval 16--17~GHz under
	TRIV with $B=190~{\rm mT}$. Panel (a) on the left-hand side
	displays on a logarithmic scale (base $10$) the $x_k$ as
	obtained from the data. Panel (b) shows analogously the
	rescaled quantities $z_k = x_k / \langle x_k \rangle$. Panel
	(c) on the right-hand side shows the distribution of $M=1200$ 
	rescaled coefficients on a logarithmic scale for six
	realizations. The dashed line is the exponential expected for
	a Gaussian-distributed $\tilde{S}$.}
	\label{fig:remove_mean}
\end{figure}

We ask: How are the $\vert\tilde{S}_k\vert^2$ distributed about their
mean values determined by the fits? In order to study the distribution
of the $\tilde{S}_k$ and of the coefficients
$x_k=\vert\tilde{S}_k\vert^2$ with good statistics, we must sample all
data points shown in Fig.~\ref{fig:goe_fft}. To this end we remove the
strong and systematic $k$ dependence by rescaling: We divide
$\tilde{S}_k$ by $\sqrt{\langle x_k \rangle}$ and $x_k$ by $\langle
x_k \rangle$ and find that the renormalized $S$-matrix elements
$\tilde{S}_k / \sqrt{\langle x_k \rangle}$ have a bivariate Gaussian
distribution both for the elastic (case shown in
Fig.~\ref{fig:S11fft_distrib}) and the inelastic one (shown in
Fig.~\ref{fig:S12fft_distrib}). The left-hand side of
Fig.~\ref{fig:remove_mean} shows that after rescaling of the $x_k$ the
logarithms of the rescaled coefficients $z_k = x_k / \langle x_k
\rangle$ scatter about zero. Moreover we find that the distribution of
the $z_k$ is stationary in $k$. By this we mean that the distribution
of the $z_k$ determined from sampling their values within some
interval of length $\delta k \ll M$ does not depend on the choice or
length of that interval. The statistical accuracy of that statement is
obviously limited by the fact that the number of data points contained
in the interval decreases with decreasing length $\delta k$.
Stationarity allows us to study the joint distribution function of all
$z_k$ obtained from $S$-matrix data that lie within a frequency
interval of length $1$ GHz. That step improves the statistical
accuracy of the result.

The right-hand side of Fig.~\ref{fig:remove_mean} shows that the
coefficients $z_k$ have an exponential distribution, as expected for
the absolute squares of variables with a bivariate Gaussian
distribution. To test this statement quantitatively we observe that
for an exponential distribution the ratio $\langle z^2_k \rangle /
\langle z_k \rangle^2$ should have the value two. For our finite
data set we define
\begin{equation}
	M_1 = \frac{1}{M} \sum_{k=0}^{M-1} z_k, \quad
	M_2 = \frac{1}{M} \sum_{k=0}^{M-1} z_k^2\, 
	\label{eqn:test_M1M2}
\end{equation}
and obtain for the variance of $M_2/M_1^2$
\begin{equation}
	\left\langle\left(\frac{M_2}{M_1^2} - 2 \right)^2 \right\rangle 
	= \frac{9}{M}\, .
	\label{eqn:test_result}
\end{equation}
Evaluation of that ratio for the data set in the range 16--17~GHz
shown in Fig.~\ref{fig:remove_mean} with 6 realizations gives
$M_2/M_1^2 = 1.97$, which is within the defined error limits
($2\pm0.09$) for the $M=6\cdot200=1200$ contributing data points. A
systematic analysis of our data ensemble for all $24$ frequency
intervals between $1$ and $25$~GHz yields 13 accepted and 11 rejected
ratios, i.e.\ 54~\% of all frequency intervals are within the
1-$\sigma$ range defined by Eq.~(\ref{eqn:test_result}). In the range
10--25~GHz the acceptance ratio increases to 80~\%. This is well
above the expected 1-$\sigma$ value of approximately $68~\%$.

\subsection{Analysis}
\label{sec:analysis}

The results displayed in Fig.~\ref{fig:remove_mean} are puzzling. The
elements of $S(f)$ are correlated over a frequency range $\Gamma$.
They do not follow a Gaussian distribution. There are non-trivial
higher-order correlations. On the other hand, there exists no
discernible correlation among the rescaled Fourier coefficients $z_k$, and
these are consistent with a bivariate Gaussian distribution for
$\tilde{S}_k / \sqrt{\langle x_k \rangle}$. How is it possible that a
non-Gaussian distribution becomes Gaussian after Fourier
transformation and rescaling?

We first show that rescaling removes binary correlations. For
simplicity, we do so for the case of a continuous frequency $f$
ranging over the entire real axis. For clarity, we distinguish the
ensemble average (indicated by an overbar) from the running average
over the spectrum of a single realization (indicated by angular
brackets).

Without rescaling, the distribution of $\tilde{S}(k)$ would obviously
not be Gaussian. Moreover, rescaling does indeed remove all
correlations between pairs of $S$-matrix elements. To see this, we
calculate the correlation function of two Fourier-transformed
$S$-matrix elements, using the translational invariance of the
two--point correlation function [$\overline{S^{\rm fl}(f_1) {S^{\rm
fl}}^\ast(f_2)} = \overline{S^{\rm fl}(f_1 + x) {S^{\rm fl}}^\ast(f_2
+ x)} = g_2(f_1 - f_2)$ for all real $x$]. That gives
\begin{equation}
	\overline{\tilde{S}(k_1) \tilde{S}^\ast(k_2)} = 2 \pi\,
	\delta(k_1 - k_2) \ \tilde{g}_2(k_1) \, .  \label{eqn:7}
\end{equation}
The Fourier transform $\tilde{g}_2 > 0$ of $g_2$ determines only the
average value of $|\tilde{S}(k)|^2$; pairs of Fourier-transformed
$S$-matrix elements with different arguments are uncorrelated. We
Fourier-transform $\tilde{S}(k) / \sqrt{\tilde{g}_2(k)}$ back to the
frequency domain and find that the correlation function of a pair of
Fourier-back-transforms is a delta function in frequency. Thus, the
binary correlation has been removed by rescaling. Put differently, we
may consider the quantities $\tilde{S}(k) / \sqrt{\tilde{g}_2(k)}$ as
Fourier transforms of $S$-matrix elements that are pairwise
uncorrelated.

Correlations of higher order (involving more than two $S$--matrix
elements) imply correlations of higher order of the elements of
$\tilde{S}^{\rm fl}(k)$ and of $\tilde{S}(k) /
\sqrt{\tilde{g}_2(k)}$. Such correlations are not removed by
rescaling. But they are made irrelevant by the way in which the
distribution of the $z_k$ is sampled. That is done by considering the
index $k$ as a label only. Any relation to the time scale originally
inherent in the Fourier transformation is lost. We simply order the
$z_k$ by size, asking how many occur in each size interval. That
yields the distribution in Fig.~\ref{fig:remove_mean}. The dependence
on $k$ is scrambled. It is not possible from that distribution to
reconstruct correlations that may have existed among its elements.

These arguments do not explain why the distribution of $\tilde{S}^{\rm
fl}(k) / \sqrt{\langle x_k \rangle}$ is Gaussian. (For that we must
resort to the law of large numbers). But they show why correlations
that are known to exist among the $S^{\rm fl}(f)$ do not prevent a
Gaussian to emerge for the distribution of the $\tilde{S}^{\rm fl}(k)
/ \sqrt{\langle x_k \rangle}$.

\subsection{Goodness-of-Fit Test}
\label{sec:gof}

To test the quality of the fit of the theoretical autocorrelation
functions in Eqs.~(\ref{eqn:CabGOE}) and Eq.~(\ref{eqn:Cab}) to the
data, we developed a goodness-of-fit test. The test applies to
uncorrelated data with an exponential distribution. As shown in
Sec.~\ref{sec:analysis} that condition is met by the rescaled
experimental Fourier coefficients $x_k / \langle x_k \rangle$. We
recall that the mean values $\langle x_k \rangle$ are determined by
fitting a small number of parameters. This renders the decision
non-trivial whether the fit is compatible with the data. As a measure
for the goodness of the fit we used the expression
\begin{equation}
	I = \sum_{k=0}^{M-1} \left( \frac{x_k}{\langle x_k \rangle} - \ln
	\frac{x_k}{\langle x_k \rangle} - 1 \right)
	\label{eqn:hlh_I}
\end{equation}
which quantifies the difference between the $M$ data points $x_k$ and
the best-fit value $\langle x_k\rangle $ for the theoretical
expressions. The quantity $I$ is non-negative and vanishes exactly if
$x_k = \langle x_k \rangle$ for all $k$. The expression
Eq.~(\ref{eqn:hlh_I}) is a generalization of the $\chi^2$ test used
for Gaussian data, see Chaps.~14 and 16 of Ref.~\cite{Harney2003}. If
$x_k / \langle x_k \rangle$ has an exponential distribution then the
distribution of $I$ is given by
\begin{equation}
	P(I) = \frac{1}{2\,\pi} \int_{-\infty}^\infty {\rm d}\alpha\,
	e^{i\,\alpha(I+M)} \left[ \frac{\Gamma\left(1+i\,\alpha
	\right)}{\left( 1+i\, \alpha \right)^{1+i\,\alpha}} \right]^M\, .
	\label{eqn:gofdistrib}
\end{equation}
For our test we approximated $P(I)$ by a chi-squared distribution
$\chi^{(M)}$ with $M$ degrees of freedom
\begin{equation}
	\chi^{(M)} (I|\bar{I}) = \frac{(M/2)^{M/2}}{\Gamma(M/2) \bar{I}} \,
	\left( \frac{I}{\bar{I}} \right)^{\frac{M}{2}-1}\, \exp \left(
	-\frac{M\, I}{2\, \bar{I}} \right)\, .
	\label{eqn:hlh_chi}
\end{equation}
Here, $\bar{I}$ is the expectation value of $I$ and is given by
\begin{equation}
	\bar{I} = M\, \gamma\, ,
	\label{eqn:barI}
\end{equation}
where $\gamma = 0.577216$ is Euler's constant. The agreement between
$P(I)$ and $\chi^{(M)}$ is better than 2~\%. The same measure $I$ was
used for a goodness-of-fit test in Ref.~\cite{Friedrich2008}.

\begin{figure}[ht]
	\includegraphics[width=8cm]{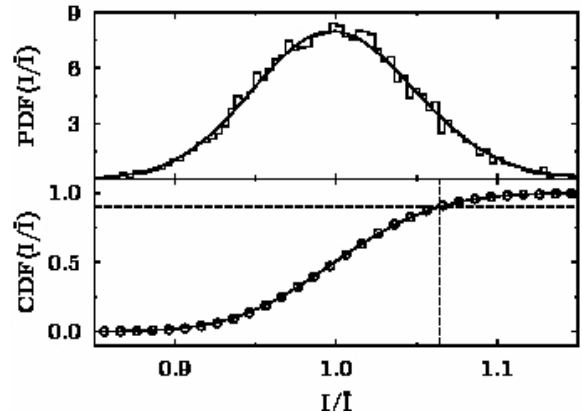}
	\caption{The probability distribution function (PDF, upper panel)
	and the cumulative distribution function (CDF, lower panel) of the
	distance measure $I/\bar{I}$ for $M=800$. The solid lines in both
	panels correspond to the analytic model Eq.~(\ref{eqn:gofdistrib}),
	the results of a Monte Carlo simulation are shown in the upper
	(lower) panel as a histogram (circles). For a threshold of $K=0.9$
	on the CDF a limit (horizontal dashed line) of $R^2_K=1.064$ is
	imposed on $I/\bar{I}$ (vertical dashed line) which may not be
	exceeded if the GOF test is to accept the model.}
	\label{fig:d_distrib_800}
\end{figure}

The test procedure is illustrated in Fig.~\ref{fig:d_distrib_800}.
Upon the definition of a certain threshold $K$ (in the Figure: $K =
0.9$) on the cumulative distribution function a limit $R^2_K$ is
obtained (in the Figure: $R^2_K = 1.064$) which may not be exceeded by
$I/\bar{I}$ if the GOF test is to accept the model. The limit is
defined by
\begin{equation}
	\frac{1}{\bar{I}} \int_0^{R_K^2} {\rm d}I\, 
	\chi^{(M)}(I|\bar{I}) = K\, .
	\label{eqn:hlh_RK2}
\end{equation}
The value of $K$ quantifies the confidence into the test in the sense
that $1-K$ is the probability to make a wrong decision by rejecting
a valid theory. We choose $K=0.9$.

\begin{figure}[ht]
	\includegraphics[width=8cm]{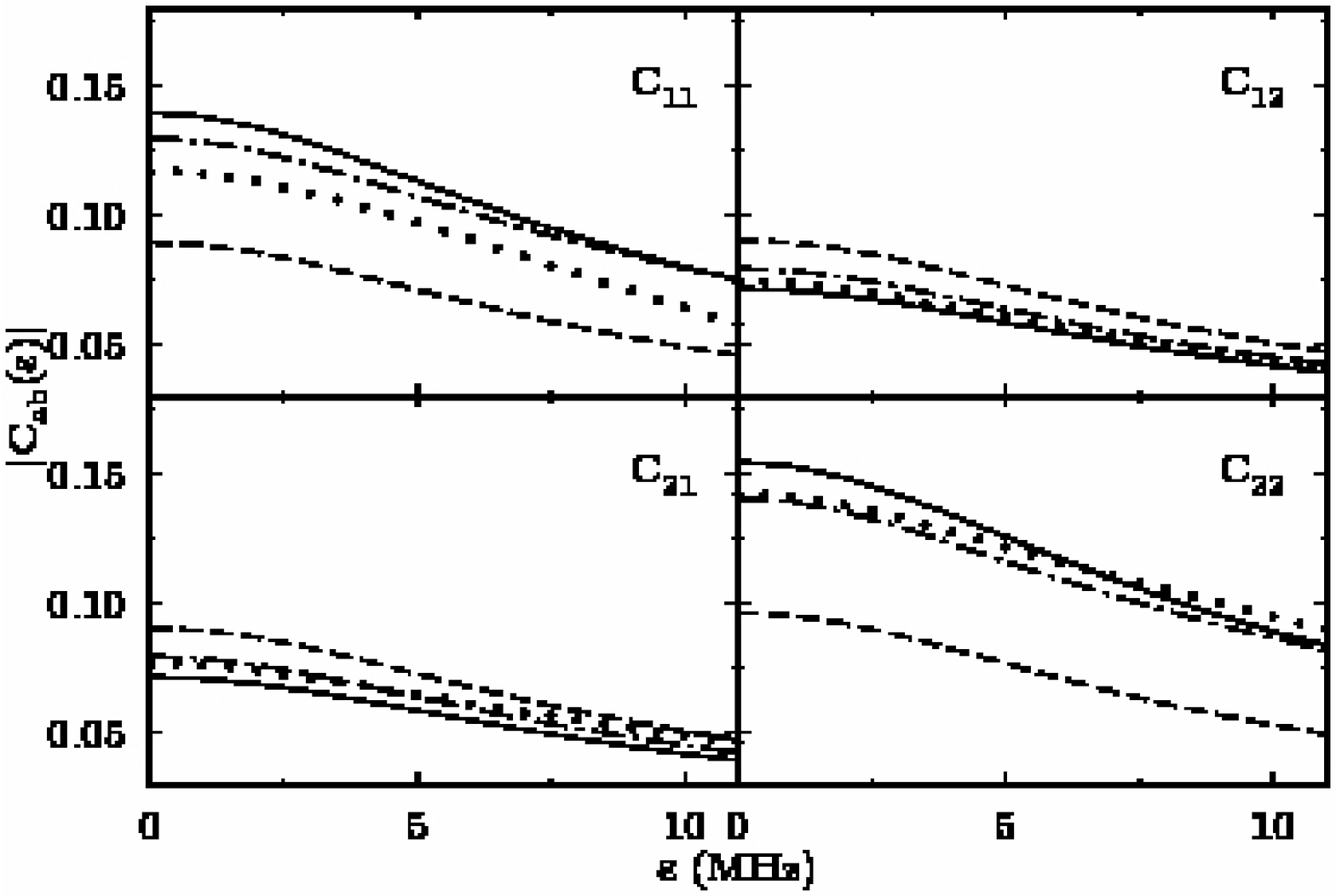}\\
	\includegraphics[width=8cm]{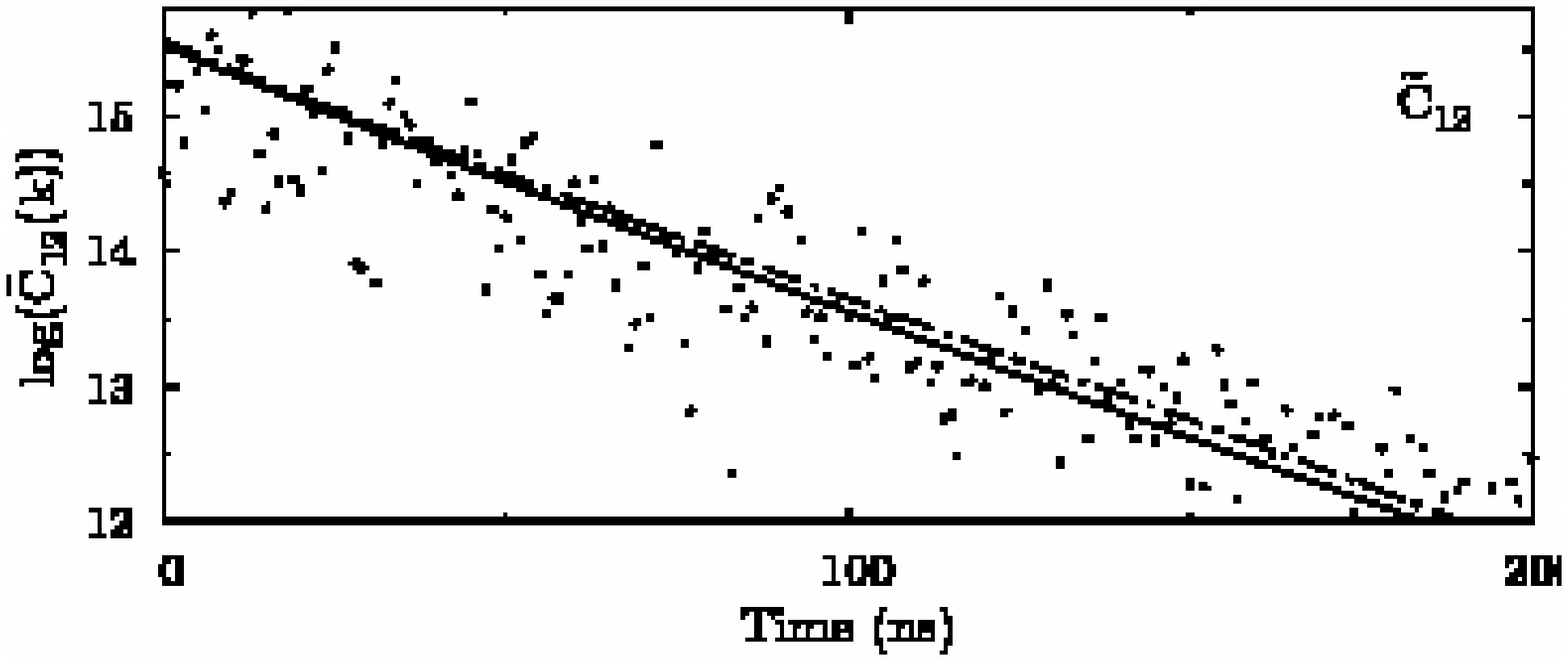}
	\caption{Top: Comparison of autocorrelation functions in the
	frequency domain for a single realization in the range 24--25~GHz,
	i.e.\ $\Gamma/d \approx 1.14$, and for $B=190~{\rm mT}$. The four
	panels display the results for $C_{11},\ C_{12},\ C_{21},\ C_{22}$.
	The discrepancy between data (dots) and VWZ (solid) or complete TRIV
	(dashed) is more pronounced than for the model for partial TRIV
	(dash-dotted) with $\xi = 0.202$. Bottom: $C_{12}$ in the time
	domain (same key, but for clarity without the result for complete
	TRIV).}
	\label{fig:correl_frq_time}
\end{figure}

In the fitting procedure the Fourier transforms of the theoretical
autocorrelation functions of the GOE (Eq.~(\ref{eqn:CabGOE})), of the
GUE (Eq.~(\ref{eqn:CabGUE})), and for the case of partial
\Ti-invariance violation (Eq.~(\ref{eqn:Cabauto})), respectively, were
fitted to the experimental Fourier coefficients. One example of such a
fit is shown in Fig.~\ref{fig:correl_frq_time}. In the upper four
panels the results are compared to the data (dots) in the frequency
domain. The lower panel shows the experimental Fourier coefficients
(dots) together with the best fits of the GOE (solid) and the partial
TRIV (dash-dotted) result.
We observe that the experimental Fourier coefficients scatter widely
around their average; the scatter is too large to directly arrive at a
conclusive decision using the GOF test described above. Therefore data
were taken from a total of six different realizations of the experiment.
Moreover, the theoretical curves lie very close to each other. This
corroborates the need of a goodness-of-fit test. This GOF test in
conjunction with the enlarged data basis leads to the decision that
within the confidence threshold $K = 0.9$ only the expression for
partial TRIV describes the data. The GOE and GUE expressions are ruled
out.

\section{Summary}
\label{sec:sum}

We have measured reflection and transmission amplitudes in chaotic
microwave billiards with two antennas. The measurements were performed 
in the regimes of isolated and
of weakly overlapping resonances. Both a \Ti-invariant system and a
system with partially violated \T invariance were investigated. The
latter was realized by placing a magnetized ferrite within the
microwave billiard. The measurements yielded the moduli and phases of
all four elements of the scattering matrix $S(f)$ in a range of 
frequencies $f$ limited by the requirement that only one vertical mode
be excited in the billiard. The frequency range was divided into
intervals of 1~GHz width. Within each interval, statistical measures
for $S$-matrix fluctuations like the Fourier transform of the
$S$-matrix autocorrelation function, the distribution of $S$-matrix
elements, or the elastic enhancement factor were determined from the
data.

We compared the results with theoretical expressions based on
random-matrix theory. For \Ti-invariant systems and for systems with
full violation of \T invariance these were given in
Refs.~\cite{Ver85,Pluhar1995,Fyodorov2005} while for systems with
partial \Ti-invariance violation they had to be calculated. This was
done by extending the existing supersymmetry approach. The parameters
of the theory are the transmission coefficients $T_1$ and $T_2$ for
the two antennas, the parameter $\tau_{\rm abs}$ describing absorption
in the billiard, and the parameter $\xi$ for the strength of
\Ti-invariance violation. Starting values for these were partly
obtained directly from the data, but final values were always
determined from fits of the RMT expressions to some of the
experimental measures.

The large data sets taken made it possible to test the theoretical
expressions with unprecedented accuracy. The outcome of these tests 
is recapitulated in Table~\ref{tab:without}. In particular, we used the
following stringent tests.

\noindent
(i) Goodness-of-fit (GOF) test. For the \Ti-invariant system, that
test accepted the fit of the theoretical result for the
Fourier-transformed autocorrelation function to the experimental data
in all 1~GHz intervals. In the case of \Ti-invariance violation, the
GOF test was applied to data fits of RMT expressions for all three
cases, i.e., the one for \Ti-invariant systems, the one for systems
with complete and the one for systems with partial \Ti-invariance
violation. The results of these fits are summarized in
Table~\ref{tab:without}. The fit is accepted in just 7 frequency
intervals for the first case, it is rejected in all but two intervals
for the second case, and it is accepted in all but one interval for
the third case. In each 1~GHz window the fitted values of $\xi$ agree
with the values determined in Sec.~\ref{fitpar}. We conclude that the
GOF test is a powerful tool to uncover the small effects of partial
\Ti-invariance violation on $S$-matrix fluctuation properties.

\begin{table}[ht]
	\centering
	\caption{Results of the GOF test for the billiard with ferrite at
	$B=190~{\rm mT}$. In each case the first row gives the lower
	boundary of the 1~GHz frequency interval used for the analysis. The
	second (third) row indicates whether the GOF test for the
	autocorrelation function for \T invariance (full violation of \T
	invariance, respectively) was accepted. This is indicated by a
	bullet. If both expressions are accepted, no conclusions can be
	drawn and the column is marked by ``$\circ$'' signs. Results
	rejected by the fit are indicated by ``$-$''. The fourth row shows
	similarly acceptance or rejection of the expression for partial
	TRIV.
	\vspace{1ex}}
	\renewcommand{\tabcolsep}{3pt}
	\begin{tabular}{l|p{9pt}p{9pt}p{9pt}p{9pt}p{9pt}p{9pt}p{9pt}p{9pt}p{9pt}p{9pt}p{9pt}p{9pt}}
		\hline
		\hline
		$f$ (GHz) & 1 & 2 & 3 & 4 & 5 & 6 & 7 & 8 & 9 & 10 & 11 & 12\\
		\hline
		no TRIV & $-$ & $\,\bullet$ & $\circ$ & $\,\bullet$ & $\circ$ & $\,\bullet$ & $\,\bullet$ & $\,\bullet$ & $\,\bullet$ & $\,\bullet$ & $\,\bullet$ & $\,\bullet$ \\
		TRIV & $-$ & $-$ & $\circ$ & $-$ & $\circ$ & $-$ & $-$ & $-$ & $-$ & $-$ & $-$ & $-$ \\
		partial & $\,\bullet$ & $\,\bullet$ & $\,\bullet$ & $\,\bullet$ & $\,\bullet$ & $\,\bullet$ & $\,\bullet$ & $\,\bullet$ & $\,\bullet$ & $\,\bullet$ & $\,\bullet$ & $\,\bullet$ \\
		\hline
		\hline
	\end{tabular}

	\vspace{1ex}

	\begin{tabular}{l|p{9pt}p{9pt}p{9pt}p{9pt}p{9pt}p{9pt}p{9pt}p{9pt}p{9pt}p{9pt}p{9pt}p{9pt}}
		\hline
		\hline
		$f$ (GHz) & 13 & 14 & 15 & 16 & 17 & 18 & 19 & 20 & 21 & 22 & 23 & 24 \\
		\hline
		no TRIV & $\,\bullet$ & $\,\bullet$ & $\,\bullet$ & $\,\bullet$ & $-$ & $-$ & $\,\bullet$ & $\,\bullet$ & $-$ & $\,\bullet$ & $\,\bullet$ & $-$ \\
		TRIV & $-$ & $-$ & $-$ & $-$ & $-$ & $-$ & $-$ & $-$ & $-$ & $-$ & $-$ & $-$ \\
		partial & $\,\bullet$ & $\,\bullet$ & $\,\bullet$ & $\,\bullet$ & $\,\bullet$ & $\,\bullet$ & $\,\bullet$ & $\,\bullet$ & $\,\bullet$ & $\,\bullet$ & $-$ & $\,\bullet$ \\
		\hline
		\hline
	\end{tabular}

	\label{tab:without}
\end{table}

\noindent
(ii) We inserted the fitted parameters into theoretical expressions
for the distribution of the diagonal $S$-matrix elements and, in the
case of \T violation, for the elastic enhancement factor. The results
agreed well with the data.  To the best of our knowledge, this is the
first time that the elastic enhancement factor as function of the
parameter $\xi$ has been investigated in such detail over such a large
frequency range.

As an additional test, we extended our measurements beyond the
frequency range where only one vertical electric mode in the resonator
is excited. When two such modes are excited, the resonator does not
simulate a quantum billiard. The fit of the theoretical result based
on random-matrix theory to the experimental data is rejected by the
GOF test. That was shown in Ref.~\cite{Friedrich2008} and is not
reproduced here. We have numerically simulated the $S$-matrix
correlations for that case under the assumption that the two modes of
vertical excitation do not interact, using for the random Hamiltonian
ensemble sets of real symmetric matrices consisting of two diagonal
blocks. Such matrices are commonly used to mimic spectral properties
of \Ti-invariant chaotic systems with some underlying symmetry. It was
shown in Ref.~\cite{Friedrich2008} that our simulations qualitatively
reproduce the experimental autocorrelation function. The failure of
the GOF test shows that our testing procedure is sensitive to the
existence of such symmetries.

We conclude that the theoretical expressions for the $S$-matrix
correlation functions, for the distribution of $S$-matrix elements,
and for the elastic enhancement factor based on random-matrix theory,
are in excellent agreement with data measured on chaotic microwave
billiards, both for the \Ti-invariant case and for the case with
partial \Ti-invariance violation. Our work constitutes the most
stringent test of the statistical theory of quantum chaotic scattering
yet done. The success in the case of partial \Ti-invariance violation
shows that the strength parameter $\xi$ can be determined reliably
from scattering data. This is important in cases where that parameter
cannot be reliably obtained theoretically from a dynamical calculation
like, for instance, the semiclassical approximation. The largest
achieved values for the \Ti-invariance violation strength parameter
$\xi$ equals 0.3. Numerical calculations show that for this value the
spectral fluctuations of the Hamiltonian $H$ for the closed resonator
defined in Eq.~(\ref{eqn:hamiltonian}) almost coincide with those of
the GUE~\cite{Bohigas1995}. We also found that for $\xi=0.4$ they do
not differ significantly from those presented in Ref.~\cite{So1995},
where the conclusion was drawn, that complete \T breaking is achieved.
However, even for $\xi =0.4$ the value of $C_{\rm cross}(0)$ is still
far from zero. This shows that $C_{\rm cross}(0)$ is a particularly
suitable measure of the strength $\xi$ of \T violation.

\begin{acknowledgments}
We would like to thank J.\ Verbaarschot for his initial help
concerning the analytic description of systems with $T$-violation.
Moreover we would like to thank D.\ Savin and Y. Fyodorov for
discussions on the distribution of the $S$-matrix elements. F.~S.\ is
grateful for the financial support from the Deutsche Telekom
Foundation. This work was supported by the DFG within SFB~634.
\end{acknowledgments}

\section*{Appendix}

For the derivation of the analytic expressions for the autocorrelation
function and the cross correlation coefficient in
Eqs.~(\ref{eqn:Cab}),~(\ref{eqn:Cabauto}) and~(\ref{eqn:Cabcross}), we
used Efetov's supersymmetry approach in the form of
Ref.~\cite{Pluhar1995}. Due to the symmetry-breaking term in
Eq.~(\ref{eqn:hamiltonian}), the integration over the Grassmann
variables which involves determining terms of highest (in our case
eighth) order in the anticommuting variables, is rather difficult. In
Ref.~\cite{Pluhar1995} the integration was done with Efetov's original
parametrization for the integration measure and a method first
developed in Ref.~\cite{Altland1992}. The starting point was the
generating functional
\begin{eqnarray}
	Z(\varepsilon)&=& {\rm Detg}^{-1} \left[\hat D + \hat
	J(\varepsilon)\right]\nonumber\\
	&=&\exp\left(-{\rm Trg}\ln\left[\hat D+\hat
	J(\varepsilon)\right]\right)\, .
	\label{eqn:Ze}
\end{eqnarray}
Here, Detg and Trg denote the graded determinant and trace as defined
in the supersymmetry formalism of Ref.~\cite{Ver85}. The inverse
propagator $\hat D$ and the matrix $\hat J$ are $4N\times 4N$
matrices. Introducing the matrix
\begin{equation}
        \Omega_{\mu\nu}^c=\pi W_{c\mu}W_{c\nu}\, .
	        \label{eqn:Omega}
		\end{equation}
we have for the inverse propagator $\hat D$
\begin{equation}
	\hat D=\left(f-\hat H\right)\mathbb{I}_4+\frac{1}{2}\epsilon
	L+i\Omega L\ .
\end{equation}
Here $\Omega =\sum_c\Omega^c$ and $L_{pp^\prime}^{\alpha\alpha^\prime}
= (-1)^{p+1}\delta_{pp^\prime} \delta^{\alpha\alpha^\prime}$ with
$p,p^\prime=1,2$ and $\alpha, \alpha^\prime=0,1$ is the diagonal
supermatrix that distinguishes between the advanced ($p=1$) and
retarded ($p=2$) parts of $\hat D$. The index $\alpha=0$ denotes the
commuting, $\alpha =1$ the anticommuting components. The quantity
$\epsilon$ equals the difference of the arguments of the $S$-matrix
elements $S_{ab}(f-\epsilon/2)$ and $S_{ab}^*(f+\epsilon/2)$. The
matrix $\hat J(\varepsilon)$ is given as
\begin{eqnarray}
	{\hat
	J_{\mu\nu}}(\{\varepsilon_{ab}^1\},\{\varepsilon_{ab}^2\}) &=&
	\pi\sum_{a,b}\sum_{j=1}^2{I}(j)W_{a\nu}
	\varepsilon_{ab}^jW_{b\mu}
	\label{eqn:J}
\end{eqnarray}
where the matrix ${I}(j)$ with entries
\begin{equation}
	I_{pp^\prime}^{\alpha\alpha^\prime}(j) = (-1)^{1+\alpha}
	\delta_{pp^\prime} \delta^{\alpha\alpha^\prime}\delta_{pj}
\end{equation}
is the projector onto the $p=j$ block. With Eq.~(\ref{eqn:Sab}) we obtain
\begin{align}
	& S_{ab}(f-\epsilon/2)S_{ab}^*(f+\epsilon/2) \nonumber \\
	& \quad = 4\, {\rm Tr}\Omega^a\, D^{-1}(f-\epsilon/2)\, \Omega^b
	\left[D^{-1}(f+\epsilon/2)\right]^\dagger\, ,
	\label{eqn:SS}
\end{align}
where ${\rm Tr}$ denotes the trace over the index $\mu $ of the
resonator modes, and it can be checked that 
\begin{equation}
	S_{ab}(f-\epsilon/2) S_{ab}^*(f+\epsilon/2) =
	\left.\frac{\partial^2}{\partial\varepsilon_{ab}^1
	\partial\varepsilon_{ba}^2} Z(\varepsilon) \right\vert_{\varepsilon
	=0}\, .
\end{equation}
Averaging over the ensemble $H$ becomes feasible when we write the
generating functional as a Gaussian superintegral
\begin{equation}
	Z(\varepsilon) = \int\mathcal{D}\Phi\, \exp
	\left(\frac{i}{2}\sum_{p,r,\alpha}
	\left\langle\overline{\Phi}_{pr}^{\alpha}, \left[\left(\bf{\hat
	D+\hat J(\varepsilon)}\right) \Phi\right]_{pr}^{\alpha}
	\right\rangle \right)
	\label{eqn:ZeG}
\end{equation}
over an eight-dimensional supervector $\Phi$. The matrices  $\hat D$
and $\hat J$ have been extended to $8N\times 8N$ supermatrices,
\begin{equation}
	{\bf\hat D}=\left(
	\begin{array}{cc}
		\hat D &0\\
		0 &\hat D^T
	\end{array}\right)\, ,
\end{equation}
and
\begin{align}
	{\bf\hat J} &= \hat J(\{\varepsilon_{ab}^{{\rm
	(S)}1}\},\{\varepsilon_{ab}^{{\rm (S)}2}\})\otimes
	\left(\begin{array}{cc}1 &0\\0 &1\end{array}\right) \nonumber \\
	& +\hat
	J(\{\varepsilon_{ab}^{{\rm (A)}1}\},\{\varepsilon_{ab}^{{\rm
	(A)}2}\})\otimes \left(\begin{array}{cc}1 &0\\0
	&-1\end{array}\right)\, ,
\end{align}	
where $\varepsilon_{ab}^{{\rm (S)}j}$ and $\varepsilon_{ab}^{{\rm
(A)}j}$ are the parts of $\varepsilon_{ab}^j$ that are symmetric and
antisymmetric in the indices $a$ and $b$, respectively. The indices
$r,r^\prime =1,2$ in Eq.~(\ref{eqn:ZeG}) arise due to the doubling of
dimension. The eight-component supervector $\Phi$ is given in terms
of the four-component supervector $\phi_p^\alpha$ and its adjoint
$\overline{\phi} = \phi^\dagger s$, where
$s_{pp^\prime}^{\alpha\alpha^\prime} = (-1)^{(\alpha +1)(1+p)}
\delta_{pp^\prime}^{\alpha\alpha^\prime}$ as
\begin{equation}
	\Phi=\left(\begin{array}{c}\phi\\ s\phi^*\end{array}\right)\, ,\
	\overline{\Phi} = \Phi^\dagger s\, .
\end{equation}
The symmetrized form of the autocorrelation function,
$\frac{1}{2}\left(S_{ab}(f-\epsilon /2)S_{ab}^\ast(f+\epsilon
/2)+S_{ba}(f-\epsilon /2)S_{ba}^\ast(f+\epsilon /2)\right)$, is
obtained by choosing the plus sign, that of the unnormalized
cross-correlation coefficient $\mathfrak{Re}\left( \left\langle
S_{ab}(f)S_{ba}^\ast(f) \right\rangle \right)$ by setting $\epsilon=0$
and choosing the minus sign in
\begin{align}
	& \frac{1}{4}\left.\frac{\partial^2}{\partial\varepsilon_{ab}^{\rm
	(S)1} \partial\varepsilon_{ba}^{\rm (S)2}}Z(\varepsilon^{\rm
	(S)},0)\right\vert_{\varepsilon =0} \nonumber \\
	&\quad \pm\frac{1}{4}\left.\frac{\partial^2}{\partial\varepsilon_{ab}^{\rm
	(A)1} \partial\varepsilon_{ba}^{\rm (A)2}}Z(0,\varepsilon^{\rm
	(A)})\right\vert_{\varepsilon =0} \ .
	\label{eqn:Gen}
\end{align}
Ensemble averaging over the Hamiltonian $H$ of the generating functional
Eq.~(\ref{eqn:ZeG}) yields
\begin{widetext}
\begin{equation}
	\left\langle{Z(\varepsilon)}\right\rangle 
        = \int\mathcal{D}\Phi e^{i\mathcal{L}(S)}
	\exp\left(i\frac{1}{2}\sum_{p,\alpha ,r}
	\left\langle\Phi_{pr}^\alpha,
	\left[\left(f\mathbb{I}+\frac{1}{2}\epsilon\hat L+i\Omega\hat L +
	\bf{\hat J}(\varepsilon)\right) \Phi\right]_{pr}^\alpha
	\right\rangle \right)\, ,
	\label{eqn:Zebar}
\end{equation}
\end{widetext}
where 
\begin{equation}
	\mathcal{L}(S) = -\frac{\lambda^2}{4N} \left[{\rm trg}S^2 +
	\frac{\pi^2\xi^2}{N}{\rm trg} \tau^3 S \tau^3 S \right]\, ,
\end{equation}
with the supermatrix $S^{\alpha ,\alpha^\prime}_{pr,p^\prime r^\prime}
= \sum_\mu\Phi_{pr}^\alpha (\mu ) \Phi_{p^\prime
r^\prime}^{\alpha^\prime}(\mu )$ and $\tau^3_{rr^\prime} =
(-1)^{r+1}\delta_{rr^\prime}$. The matrix $\hat L$ is
eight-dimensional. It results from the doubling of dimension, $\hat
L_{rr^\prime}=L\delta_{rr^\prime}$. The quartic dependence of
$\mathcal{L}(S)$ on $\Phi$ is eliminated with help of a
Hubbard-Stratonovich transformation. After expanding the resulting
exponent in the large $N$ limit in the small quantities $\epsilon$ and
$\frac{\pi^2\xi^2}{N}$ the remaining integral reads
\begin{equation}
	\left\langle{Z(\varepsilon)}\right\rangle = \int\mathcal{D} Q
	e^{i\mathcal{L}_{\rm eff}(Q)} e^{i\mathcal{L}_{\rm
	src}(Q;{\bf \hat J})}\, ,
\end{equation}
where $\mathcal{L}_{\rm eff}(Q)=\mathcal{L}_{\rm
free}(Q)+\mathcal{L}_{\rm ch}(Q)$ and with $X_c=\pi Nv_c^2$ (c.f.\ 
Eq.~(\ref{eqn:sumr}))
\begin{align}
	i\mathcal{L}_{\rm free} &= -\frac{\pi^2\xi^2}{4} {\rm trg} \tau^3
	Q\tau^3 Q + i\frac{N\epsilon}{4\lambda} {\rm trg}\hat L Q
	\nonumber\\ 
	i\mathcal{L}_{\rm ch} &= -\frac{1}{2} \sum_c{\rm trg}
	\ln\left(\mathbb{I} + \frac{X_c}{\lambda} \hat L Q \right)
	\nonumber\\
	i\mathcal{L}_{\rm src} &= -\frac{1}{2} {\rm Trg} \ln
	\left(\mathbb{I} + \frac{Q}{i\lambda} \left(\mathbb{I}+\Omega \hat L
	\frac{Q}{\lambda}\right)^{-1} {\bf\hat J}(\varepsilon) \right)\,.
	\label{eqn:Lagr}
\end{align}
Here, trg denotes the supertrace over the indices $p,\alpha, r$, and
Trg includes the summation over the level index $\mu$. The $8\times 8$
supermatrix $Q$ contains the commuting and the Grassmannian
integration variables. For the integration Efetov's original
parametrization of the $Q$ matrix was used.

Before performing the integration
$\left\langle{Z(\varepsilon)}\right\rangle$ needs to be differentiated
twice with respect to $\varepsilon_{ab}^{\rm j}$ and
$\varepsilon_{ba}^{\rm j}$ at $\varepsilon_{ab}^{\rm j}=0$ and
$\varepsilon_{ba}^{\rm j}=0$ for $j=1,2$, c.f.\ Eq.~(\ref{eqn:Gen}).
Only terms of second order in $\varepsilon_{ab}^{\rm j}$ and
$\varepsilon_{ba}^{\rm j}$ survive. Since ${\bf\hat J}(\varepsilon)$
is linear in $\varepsilon_{ab}$, these are obtained from the second
order term in ${\bf\hat J}$ of $\exp\left(-\frac{1}{2}{\rm
Trg}\ln\left(\mathbb{I}+\hat M{\bf\hat J}\right)\right)$ which equals
$\frac{1}{2}\left[\left(-\frac{1}{2}{\rm Trg}\left(\hat M{\bf\hat
J}\right)\right)^2 +\frac{1}{2}{\rm Trg}\left(\hat M{\bf\hat J}\hat
M{\bf\hat J}\right)\right]$ with $\hat M=\frac{Q}
{i\lambda}\left(\mathbb{I}+\Omega\hat L\frac{Q}{\lambda}\right)^{-1}$.
Then we finally obtain (see Eq.~(\ref{eqn:Gen}))
\begin{widetext}
	\begin{align}
	&\frac{1}{4} \left. \frac{\partial^2}{\partial
	\varepsilon_{ab}^{\rm (S)1} \partial \varepsilon_{ba}^{\rm (S)2}}
	Z(\varepsilon^{\rm (S)},0) \right\vert_{\varepsilon = 0} \pm
	\frac{1}{4} \left. \frac{\partial^2}{\partial \varepsilon_{ab}^{\rm
	(A)1} \partial \varepsilon_{ba}^{\rm (A)2}}
	Z(0,\varepsilon^{\rm(A)}) \right\vert_{\varepsilon = 0} \nonumber\\ 
	&\quad = -\frac{1}{2} \int\mathcal{D} Q \exp\left( - \frac{1}{2}
	\sum_c{\rm trg} \ln \left(\mathbb{I} + \frac{X_c}{\lambda} \hat LQ
	\right) - \frac{\pi^2 \xi^2}{4} {\rm trg} \left(\tau^3 Q \tau^3 Q
	\right) + i \frac{N\epsilon}{4\lambda} {\rm trg} \hat L Q \right)
	\nonumber\\ 
	&\qquad \times \sum_{j=1}^2 \left[{\rm trg} \left(Q
	\frac{\frac{X_a}{\lambda}}{\left( \mathbb{I} +
	\frac{X_a}{\lambda}\hat L Q \right)} I_j(1)
	\frac{\frac{X_b}{\lambda}}{\left( \mathbb{I} + \frac{X_b}{\lambda}
	\hat L Q \right)} I_j(2) \right) \left( 1 + \delta_{ab}(1\pm 1)
	\frac{(1\mp (-1)^j)}{2} \right) \right. \nonumber\\ 
	&\qquad + \left.{\rm trg} \left( Q \frac{\frac{X_a}{\lambda}}{\left(
	\mathbb{I} + \frac{X_a}{\lambda} \hat L Q \right)} I_1(1) \right)
	{\rm trg} \left( Q \frac{\frac{X_b}{\lambda}}{\left( \mathbb{I} +
	\frac{X_b}{\lambda} \hat L Q \right)} I_1(2) \right)
	\delta_{ab}(1\pm 1) + a \leftrightarrow b \right]\, .
\end{align}
\end{widetext}
For the integration over the Grassmann variables, we proceeded as in
appendix B of Ref.~\cite{Pluhar1995}. The result is a threefold
integral. For arbitrary values of the transmission coefficients $T_c$,
the expression for the autocorrelation function for the case $a \ne b$
is obtained from that given in Eq.~(2) of Ref.~\cite{Gerland1996} by
including in the integrand the factor
$\exp\left(i\frac{\epsilon\pi}{d} \left(\lambda_1\lambda_2 -
\lambda_0\right)\right)$ arising for non-vanishing $\epsilon$ from the
last term in the first line of Eq.~(\ref{eqn:Lagr}). Here, $\lambda_0,
\lambda_1, \lambda_2$ are the integration variables. It is
straightforward to compute the autocorrelation function for the case
$a=b$ by proceeding as in appendix B of \cite{Pluhar1995}. The result
for the cross-correlation coefficient is obtained by multiplying the
second and the third rectangular bracket in Eq.~(2) of
\cite{Gerland1996} by $(-1)$. All this yields for the autocorrelation
function and the cross-correlation coefficient the expressions given
in Eqs.~(\ref{eqn:Cab}),~(\ref{eqn:Cabauto}) and~(\ref{eqn:Cabcross}).


\end{document}